\documentclass[aps,10pt,superscriptaddress,reprint,showpacs]{revtex4}

\usepackage{times}
\usepackage{amsfonts}
\usepackage{amsmath}
\usepackage{amssymb}
\usepackage{epsfig}
\usepackage{graphicx}

\begin{document}

\title{Resistive MHD model for cylindrical plasma expansion in a magnetic field}
\author{H. B. Nersisyan}
\email{hrachya@irphe.am}
\affiliation{Theoretical Physics Division, Institute of Radiophysics and Electronics,
0203 Ashtarak, Armenia}
\affiliation{Centre of Strong Fields Physics, Yerevan State University, Alex Manoogian
str. 1, 0025 Yerevan, Armenia}

\author{K. A. Sargsyan}
\affiliation{Theoretical Physics Division, Institute of Radiophysics and Electronics,
0203 Ashtarak, Armenia}

\author{D. A. Osipyan}
\affiliation{Theoretical Physics Division, Institute of Radiophysics and Electronics,
0203 Ashtarak, Armenia}

\author{M. V. Sargsyan}
\affiliation{Theoretical Physics Division, Institute of Radiophysics and Electronics,
0203 Ashtarak, Armenia}

\date{\today }

\begin{abstract}
The study of hot plasma expansion in a magnetic field is of interest for many laboratory and
astrophysical applications. In this paper, an exactly solvable analytical model is proposed
for an expanding resistive plasma in an external magnetic field in the regime in which the
magnetic field does not perturb the plasma motion. The model is based on a class of exact
solutions for the purely radial expansion of the plasma in the absence of a magnetic field.
This approximation permits the reduction of the electromagnetic problem to consideration of
a diffusion equation for the magnetic field. Explicit solutions are derived for a resistive
cylindrical plasma expanding into a uniform ambient magnetic field. Some numerical examples
related to the laser-produced plasma experiments are presented.
\end{abstract}

\pacs{52.30.-q, 03.50.De, 41.20.Gz, 52.65.Kj}
\maketitle

\section{Introduction}
\label{sec:1}

The problem of hot plasma expansion into a vacuum or into a background plasma in the presence of
an external magnetic field has been discussed in the analysis of many astrophysical and laboratory
applications (see, e.g., Refs.~\cite{zak03,win05,col11,osi03,ner09} and references therein).
In particular, such process is a topic of intense interest across a wide variety of disciplines,
with applications to solar \cite{low01} and magnetospheric \cite{hae86,ber92} physics,
astrophysics \cite{rem00}, and pellet injection for tokamak refueling \cite{str00}.

In this paper we consider analytically a resistive cylindrical plasma expanding into vacuum in the
presence of an external magnetic field. Similar problem has been treated previously in
Refs.~\cite{par80,ber68} but for a spherical plasma expansion. Of the vast literature on the
theory of plasma expanding
into a vacuum or into a background plasma, Refs.~\cite{kat61,rai63,dit00,ner06,ner10,ner11,ner12,and80,joh82,tak11,ber68,par80}
(see also references therein) illustrate various aspects and approaches. In Refs.~\cite{kat61,rai63,dit00,ner06,ner10,ner12}
plasma has been considered as a highly conducting medium with zero magnetic field inside. From the
point of view of electrodynamics, it is similar to the expansion of a superconductor in a magnetic
field. An exact analytic solution for a uniformly expanding, highly conducting plasma sphere in an
external uniform and constant magnetic field has been obtained in Ref.~\cite{kat61}. The
nonrelativistic limit of this theory has been used by Raizer \cite{rai63} to analyze the energy balance
(energy emission and transformation) during plasma expansion. A similar problem has been considered
in Ref.~\cite{dit00} in a one-dimensional (1D) geometry for a plasma layer. In our recent papers,
we obtained an exact analytic solution for the uniform relativistic expansion of a highly conducting
plasma sphere \cite{ner06,ner12} or cylinder \cite{ner10} in the presence of a dipole or homogeneous
magnetic field, respectively.

The mentioned treatments \cite{kat61,rai63,dit00,ner06,ner10,ner12} were obtained assuming a somewhat
idealized situation: uniform expansion, infinite electrical conductivity of a plasma, etc. More realistic
models for plasma expansion taking into account the deceleration (or acceleration) of the plasma boundary
have been developed, for instance, in Refs.~\cite{ner11,and80,joh82,tak11} (see also references
therein) for spherical \cite{ner11}, planar (1D) \cite{and80,joh82} and cylindrical \cite{tak11} expansions
employing ideal magnetohydrodynamic (MHD) equations. However, it should be noted that the ideal MHD may
not be justified in some experimental situations where the typical parameters are such that the plasma
resistivity is not negligible \cite{zak03,col11,par80,ber68}. In this case the coupling of the magnetic
field with the plasma motion, determined by the magnetic Reynolds number, should result in a distortion
and diffusion of the field across the expanding plasma \cite{par80,ber68}. We present here calculations of
the electromagnetic field configuration in the stages preceding significant deceleration of the plasma
and in the regime in which the magnetic field does not perturb the purely radial motion of the cylindrical
plasma. The latter assumption is valid at large initial ratios of plasma energy density to magnetic field
energy density.

\section{Resistive MHD model}
\label{sec:2}

Usually, the motion of the expanding plasma boundary is approximated as the
motion with constant velocity (uniform expansion). In the present study, a
quantitative analysis of plasma dynamics is developed on the basis of a
cylindrical model. Within the scope of this analysis, the nonuniform plasma
expansion process is examined. We consider a resistive collision-dominated
magnetized plasma expanding into vacuum in the presence of a uniform and
constant magnetic field. The relevant equations governing the expansion are
those of resistive MHD \cite{boy03}, assuming that the characteristic
length scales for plasma flow are much larger than the Debye length and
Larmor radius of the ions. Thus
\begin{eqnarray}
&&\frac{\partial \rho }{\partial t}+\boldsymbol{\nabla }\cdot \left( \rho \mathbf{u}\right) =0,  \nonumber \\
&&\rho \left[ \frac{\partial \mathbf{u}}{\partial t}+\left( \mathbf{u}\cdot
\boldsymbol{\nabla }\right) \mathbf{u}\right] +\boldsymbol{\nabla }p=\frac{1}{4\pi}
\left[ \boldsymbol{\nabla }\times \mathbf{H}\right] \times \mathbf{H} ,  \nonumber  \\
&&\frac{\partial \mathbf{H}}{\partial t}=\boldsymbol{\nabla }\times \left[
\mathbf{u}\times \mathbf{H}\right] +D\nabla ^{2}\mathbf{H} ,  \label{eq:a1}
\end{eqnarray}%
with $\boldsymbol{\nabla }\cdot \mathbf{H}=0$, where $\mathbf{H}$ is the
magnetic field, $D=c^{2}/4\pi \sigma $ is the diffusion coefficient, $\rho $,
$\mathbf{u}$, $p$ and $\sigma $ are the mass density, the velocity, the
pressure and the electrical conductivity of the plasma, respectively. In
this paper we assume an isotropic and homogeneous electrical conductivity
(and hence the diffusion coefficient $D$) of the plasma $\sigma $. The
equations above must be accompanied by the equation of state and the
equation for entropy. Using the thermodynamic relation between entropy,
pressure, and internal energy as well as Eq.~\eqref{eq:a1} the equation for
pressure reads \cite{boy03}
\begin{equation}
\frac{\partial p}{\partial t}+(\mathbf{u}\cdot \boldsymbol{\nabla })p+\gamma
p\left( \boldsymbol{\nabla }\cdot \mathbf{u}\right) =(\gamma -1)\frac{j^{2}}{\sigma }.
\label{eq:a2}
\end{equation}%
Here $\gamma $ is the ratio of the specific heats, $\mathbf{j}$ is the
current density in a plasma which after the elimination of some unimportant
terms from the generalized Ohm's law \cite{boy03} is reduced to the form
\begin{equation}
\mathbf{j}=\sigma \left( \mathbf{E}+\frac{1}{c}[\mathbf{u}\times \mathbf{H}]\right) ,
\label{eq:ohm}
\end{equation}%
where $\mathbf{E}$ is the electric field. It is convenient to introduce the
vector potential $\mathbf{A}$. Within the scope of the present study the
free charge density is absent and a suitable gauge $\boldsymbol{\nabla }%
\cdot \mathbf{A}=0$ allows the electric and magnetic fields to be determined
from the vector potential $\mathbf{A}$,
\begin{equation}
\mathbf{E}=-\frac{1}{c}\frac{\partial \mathbf{A}}{\partial t},\qquad \mathbf{%
H}=\boldsymbol{\nabla }\times \mathbf{A} .
\label{eq:a3}
\end{equation}%
Then from the last expression in Eq.~\eqref{eq:a1} one can derive a similar
equation for the vector potential $\mathbf{A}$
\begin{equation}
\frac{\partial \mathbf{A}}{\partial t}={\mathbf{u}}\times \left[ \boldsymbol{%
\nabla }\times \mathbf{A}\right] +D\nabla ^{2}\mathbf{A} .
\label{eq:diff}
\end{equation}

In deriving Eq.~\eqref{eq:diff} we have neglected the displacement current
which is justified for the nonrelativistic expansion of the plasma. More
specifically this approximation is valid at $4\pi \sigma \tau _{\mathrm{D}%
}\gg 1$, where $\tau _{\mathrm{D}}=4\pi \sigma R^{2}/c^{2}$ is the
characteristic diffusion time of the magnetic field and $R$ is the
characteristic size of the system, taken here as the radius of the
cylindrical plasma. Alternatively, this inequality implies that $c\tau _{%
\mathrm{D}}$ is much larger than the plasma radius, $c\tau _{\mathrm{D}}\gg
R$, which is well justified for the nonrelativistic expansion velocities.

The neglect of the Hall current in Eq.~\eqref{eq:ohm} and the assumption of
a scalar electrical conductivity for the magnetized plasma is justified when
the characteristic time for the Coulomb collisions $\nu _{c}^{-1}$ (where
$\nu _{c}$ is the collision frequency) is much less than the cyclotron
period of the electron.
Therefore, in this highly collisional regime the conductivity $\sigma $ is essentially a
function of plasma temperature alone \cite{spi62}. Hereafter it is assumed that the
plasma temperature and the conductivity $\sigma (t)$ are uniform and are the
functions only of time.

The system of Eqs.~\eqref{eq:a1}-\eqref{eq:diff} can be used for
determination of the dynamics of the expanding plasma as well as for the
investigation of the evolution of the induced electromagnetic fields. For
further simplification of this system we note that as long as the Lorentz
force density $\frac{1}{c}[\mathbf{j}\times \mathbf{H}]$ and the Joule
dissipation $j^{2}/\sigma $ terms are negligible compared with the
hydrodynamic terms in the left hand side of Eqs.~\eqref{eq:a1} and %
\eqref{eq:a2}, respectively, the plasma motion is a free radial expansion.
For many realistic situations with laser-produced plasmas the free expansion
is realized when the kinetic energy density of the expanding plasma is
greater than the magnetic field energy density \cite{zak03} (high-beta plasma),
$\rho u^{2}/2>H_{0}^{2}/8\pi $, where $\mathbf{H}_{0}$ is the strength of the initial unperturbed magnetic field.
This is a necessary condition that the expansion remains radial and
cylindrical. It has been shown \cite{ner11} that the system of
Eqs.~\eqref{eq:a1} and \eqref{eq:a2} allows in this case the
self-similar solutions for the quantities $\rho $, $\mathbf{u}$, and $p$
which are realized under specified initial conditions. These
solutions are characterized by a radial velocity distribution linearly
dependent on the radial coordinate $r$. At $r\leqslant R(t)$, $u_{r}(r,t)=r [\dot{R}(t)/R(t)]$,
where $R(t)$ and $\dot{R}(t)$ are the radius and the velocity of the plasma boundary.
In addition, the velocity $u_{r}(r,t) $
vanishes at $r>R(t)$, $u_{r}(r,t)=0$. The self-similar solutions for the
density $\rho $ and the pressure $p$ as well as the criterion of the
violation of the free expansion solutions are discussed in Ref.~\cite{ner11}.
However, we would like to
emphasize that since the hydrodynamic terms are reduced rapidly as the
plasma expands and the Joule heating increases the plasma temperature and
the electrical conductivity the electromagnetic terms in the right hand side
of Eqs.~\eqref{eq:a1} and \eqref{eq:a2} will no longer be negligible at the
final stage of the plasma expansion when plasma may fully be stopped and
deformed by the magnetic field pressure. As mentioned above, the average
plasma pressure $\overline{p}$ is strongly reduced compared to the magnetic
pressure and the model of the purely radial expansion clearly becomes
invalid in this case. Nevertheless, if the critical time interval $\Delta t$,
where $\overline{p}<H^{2}/8\pi $, is much smaller than the typical time
scale of the plasma flow (up to the full stop), the contribution of this
interval to the overall plasma dynamics is negligible and use of the radial
expansion model is justified.

In the next sections we will use the profile of the plasma radial flow velocity $u_{r}(r,t)=r [\dot{R}(t)/R(t)]$
together with Eq.~\eqref{eq:diff} to investigate the electromagnetic field configuration generated by the expanding
cylindrical plasma.

\section{Solution of the moving boundary and initial value problem}
\label{sec:3}

In this section we consider the moving boundary problem of the plasma
cylinder expansion in the vacuum in the presence of the constant and
homogeneous magnetic field $\mathbf{H}_{0}$. Consider a cylindrical region
of space with radius $r=R(t)$ at the time $t$ containing a neutral plasma
which has expanded at $t=0$ (with $R(0)=R_{0}$) to its present state from a
cylindrical source with radius $R_{0}$ located around $r=0$. We assume that
at any time $t$ the plasma cylinder is unbounded in $z$ direction (i.e. the
plasma cylinder is located at $-\infty <z<\infty $). To solve the boundary
problem we introduce the cylindrical coordinate system ($r$, $\varphi $, $z$)
with the $z$-axis along the plasma cylinder symmetry axis and the
azimuthal angle $\varphi $ is counted from the plane ($xz$-plane) containing
the vector of the unperturbed magnetic field $\mathbf{H}_{0}$. The angle $%
\theta $ between the vector $\mathbf{H}_{0}$ and the $z$-axis is arbitrary.

As the cylindrical plasma expands it both perturbs the external
magnetic field and generates an electric field. We shall obtain an analytic
solution of the electromagnetic field configuration.
We consider the case of the purely radial expansion of the plasma cylinder
with an arbitrary (but nonrelativistic) expansion velocity $\dot{R}(t)$.
Having in mind the symmetry of the unperturbed magnetic field
and the fact that the electromagnetic fields do not depend on the coordinate
$z$ it is sufficient to choose the vector potential in the form $A_{r}=0$,
\begin{equation}
H_{0\parallel }W(r,t)=\frac{\partial }{\partial r} (rA_{\varphi }) ,
\quad  A_{z} =H_{0\perp }\Psi (r,t)\sin \varphi ,
\label{eq:1}
\end{equation}%
where $W(r,t)$ and $\Psi (r,t)$ are some unknown functions. From symmetry
considerations the functions $W(r,t)$ and $\Psi (r,t)$ are independent on
the cylindrical coordinate $\varphi $. Here $H_{0\bot }$ and $H_{0\parallel} $
are the components of the unperturbed magnetic field $\mathbf{H}_{0}$
transverse and parallel to the $z$-axis, respectively. The components of the
electromagnetic field are expressed by these functions as $H_{z}=H_{0\parallel } (W/r)$,
\begin{eqnarray}
&&H_{r}=H_{0\perp }\frac{\Psi }{r}\cos \varphi , \quad H_{\varphi }=-H_{0\perp }%
\frac{\partial \Psi }{\partial r}\sin \varphi ,  \label{eq:2} \\
&&E_{\varphi }=-\frac{1}{c}\frac{\partial A_{\varphi }}{\partial t}, \quad
E_{z}=-\frac{1}{c}H_{0\perp }\frac{\partial \Psi }{\partial t}\sin \varphi ,
\label{eq:3}
\end{eqnarray}%
and $E_{r}=0$. The equation for the vector potential $\mathbf{A}(\mathbf{r},t)$ inside the plasma
cylinder ($r\leqslant R(t)$) is obtained from the MHD equation for the magnetic field diffusion,
Eq.~\eqref{eq:diff}, which for the unknown functions $\Psi (r,t)$ and $W(r,t)$ yields a system of
equations
\begin{eqnarray}
&&\frac{\partial \Psi }{\partial t}+u_{r}\frac{\partial \Psi }{\partial r}%
=D\left( \frac{\partial ^{2}\Psi }{\partial r^{2}}+\frac{1}{r}\frac{\partial
\Psi }{\partial r}-\frac{\Psi }{r^{2}}\right) ,  \label{eq:4} \\
&&\frac{\partial W}{\partial t}+\frac{\partial }{\partial r}\left(
u_{r}W\right) =D\left( \frac{\partial ^{2}W}{\partial r^{2}}-\frac{1}{r}%
\frac{\partial W}{\partial r}+\frac{W}{r^{2}}\right) .  \label{eq:5}
\end{eqnarray}%
Here $u_{r}(r,t)=r[\dot{R}(t)/R(t)]$. In the
vacuum surrounding the plasma cylinder ($r\geqslant R(t)$), the magnetic
field is determined by the Maxwell equation $\boldsymbol{\nabla }\times
\mathbf{H}=\frac{4\pi }{c}\mathbf{j}$ with $\mathbf{j}=0 $ which for the
functions $\Psi (r,t)$ and $W(r,t)$ become
\begin{eqnarray}
&&\frac{\partial ^{2}\Psi }{\partial r^{2}}+\frac{1}{r}\frac{\partial \Psi }{%
\partial r}-\frac{\Psi }{r^{2}}=0,  \label{eq:6} \\
&&\frac{\partial ^{2}W}{\partial r^{2}}-\frac{1}{r}\frac{\partial W}{%
\partial r}+\frac{W}{r^{2}}=0.  \label{eq:7}
\end{eqnarray}%
In the plasma the magnetic field is the solution of the diffusion equation
with a diffusion coefficient $D(t)=c^{2}/4\pi \sigma (t)$. The plasma
conductivity $\sigma (t)$ is essentially a function of plasma temperature
alone. If it is assumed that the plasma temperature is uniform, this
coefficient is a function only of time and Eqs.~\eqref{eq:4}-\eqref{eq:7}
may be solved by the method of separation of variables.

The system of equations~\eqref{eq:4}-\eqref{eq:7} is to be solved in the
internal ($r\leqslant R(t)$) and external ($r\geqslant R(t)$) regions
subject to the boundary and initial conditions. Since the plasma under
consideration has finite electrical conductivity, there are no surface
currents at the plasma-vacuum boundary and $\mathbf{H}$ must be continuous
at the moving boundary. The continuity of the magnetic field $\mathbf{H}$
requires that $\Psi (r,t)$, $W(r,t)$ and the radial derivative $\partial
\Psi /\partial r$ be continuous at the expanding plasma surface. In
addition we require that as $r\to \infty $, the magnetic field
$\mathbf{H}(r,\varphi ,t)$ be time-independent and asymptotically approach
the uniform magnetic field $\mathbf{H}_{0}$. This is equivalent to the
boundary conditions $\Psi (r,t)=W(r,t)=r$ at $r\to \infty $.

The initial conditions are at $t=0$. In principle two distinct sets of
initial conditions could be considered \cite{ber68}.

(i) In the case of the poorly conducting plasma the initial conditions are
imposed for arbitrary $r$:
\begin{equation}
\Psi \left( r,0\right) =W\left( r,0\right) =r .
\label{eq:8}
\end{equation}

(ii) In the case of the perfectly conducting plasma the initial conditions
are imposed separately for the domains inside ($r\leqslant R_{0}$ with
$R_{0}=R(0)$) and outside ($r\geqslant R_{0}$) the plasma cylinder \cite{ner10}:
\begin{eqnarray}
&&\Psi \left( r,0\right) =W\left( r,0\right) =0, \quad r\leqslant R_{0} ,  \nonumber \\
&&\Psi \left( r,0\right) =r-\frac{R_{0}^{2}}{r}, \quad W\left( r,0\right)
=r , \quad r\geqslant R_{0} .  \label{eq:9}
\end{eqnarray}%
In the first case, Eq.~\eqref{eq:8}, the plasma is initially called and poorly conducting so that the
external magnetic field is completely penetrated inside the plasma. At the other extreme case (ii), the
initial conditions~\eqref{eq:9} imply that initially the plasma is highly conducting so that the magnetic
field is completely excluded from the initial plasma volume. We consider below the initial value problem
(ii). The extension of the obtained solution to the case (i) is straightforward.

At $r\geqslant R(t)$ we look for the solutions of Eqs.~\eqref{eq:6} and \eqref{eq:7} for the functions
$\Psi (r,t)$ and $W(r,t)$ in the form $\sim r^{\alpha }$, where $\alpha $ is some numerical constant.
Therefore, taking into account the boundary condition at $r\to \infty $ the full solution in the domain
outside the plasma cylinder is given by
\begin{eqnarray}
&&\Psi (r,t)=r-C(t)\frac{R_{0}^{2}}{r},  \label{eq:10} \\
&&W(r,t)=r\left[ 1+C_{1}(t)\ln \frac{r}{R(t)}\right] ,  \label{eq:11}
\end{eqnarray}%
where $C(t)$ and $C_{1}(t)$ are the arbitrary functions of time with the initial conditions $C(0)=1$ and
$C_{1}(0)=0$. However, since the magnetic field should be finite at $r\to \infty $ we set $C_{1}(t)=0$. At
$r\geqslant R(t)$ this gives the final solution $W(r,t)=r$ for the function $W(r,t)$.

For the class of separable solutions the motion is such that, for a given
element of plasma, the quantity $\xi =r/R(t)$ is a constant of the motion.
The solution of Eqs.~\eqref{eq:4} and \eqref{eq:5} inside the plasma
cylinder (i.e. at $r\leqslant R(t)$) is then facilitated by the
representation of the functions $\Psi (r,t)$ and $W(r,t)$ in the form
\begin{eqnarray}
&&\Psi (r,t)=r+\sum_{n=1}^{\infty }a_{n}T_{n}(t)\Phi _{n}(\xi ),
\label{eq:12} \\
&&W(r,t)=r+\frac{1}{R(t)}\sum_{n=1}^{\infty }b_{n}U_{n}(t)\Theta _{n}(\xi ),
\label{eq:13}
\end{eqnarray}%
where $T_{n}(t)$, $\Phi _{n}(\xi )$, $U_{n}(t)$, and $\Theta _{n}(\xi )$ are some unknown functions,
and $a_{n}$ and $b_{n}$ are the unknown expansion coefficients. Next inserting these expansions into
Eqs.~\eqref{eq:4} and \eqref{eq:5} for $\Phi _{n}(\xi )$ and $\Theta _{n}(\xi )$ one arrives at the
ordinary differential equations for the cylindrical functions \cite{bat53}. We choose only the regular
solutions of the obtained equations finite at the origin $r=0$ (or at $\xi =0 $). Thus
\begin{equation}
\Phi _{n}(\xi )=A_{n}J_{1}(\lambda _{n}\xi ), \quad \Theta _{n}(\xi
)=B_{n}\xi J_{0}(\kappa _{n}\xi ) .
\label{eq:14}
\end{equation}%
Here $A_{n}$ and $B_{n}$ are the integration constants, $\lambda _{n}$
and $\kappa _{n}$ are some arbitrary parameters (depending only on $n$),
arising due to the separation of the variables, and $J_{0}$ and $J_{1}$ are
the Bessel functions of the first kind.

In the same way for the time-dependent functions $T_{n}(t)$ and $U_{n}(t)$
we obtain the following set of the ordinary differential equations:
\begin{eqnarray}
&&\dot{T}_{n}(t)+\lambda _{n}^{2}\dot{\vartheta}(t)T_{n}(t)=\dot{R}(t),
\label{eq:15} \\
&&\dot{U}_{n}(t)+\kappa _{n}^{2}\dot{\vartheta}(t)U_{n}(t)=2R(t)\dot{R}(t),
\label{eq:16}
\end{eqnarray}%
where
\begin{equation}
\vartheta (t)=\int_{0}^{t}\frac{D(\tau )}{R^{2}(\tau )}d\tau .
\label{eq:17}
\end{equation}%
The general solutions of the first order differential equations~\eqref{eq:15}
and \eqref{eq:16} can be represented in the form
\begin{eqnarray}
&&T_{n}(t)=e^{-\lambda _{n}^{2}\vartheta (t)} \bigg[ t_{0}+\int_{0}^{t}e^{%
\lambda _{n}^{2}\vartheta (\tau )}\dot{R}(\tau )d\tau \bigg] , \label{eq:18} \\
&&U_{n}(t)=e^{-\kappa _{n}^{2}\vartheta (t)}\bigg[ u_{0} +2\int_{0}^{t}e^{%
\kappa _{n}^{2}\vartheta (\tau )}R(\tau )\dot{R}(\tau )d\tau \bigg] .
\label{eq:19}
\end{eqnarray}%
Here $t_{0} =T_{n}(0)$ and $u_{0} =U_{n}(0)$ are the initial values of $T_{n}(t)$ and $U_{n}(t)$,
respectively, to be determined by imposing the initial conditions. In addition, it should be
emphasized that the expansions given by Eqs.~\eqref{eq:12} and \eqref{eq:13} are the solutions of
the system of Eqs.~\eqref{eq:4} and \eqref{eq:5} only if the functions $\Phi _{n}(\xi )$ and
$\Theta _{n}(\xi )$ satisfy at arbitrary $0\leqslant \xi \leqslant 1$ the equations
\begin{equation}
\sum_{n=1}^{\infty }a_{n}\Phi _{n}(\xi )=\sum_{n=1}^{\infty }b_{n}\Theta
_{n}(\xi )=-\xi .
\label{eq:20}
\end{equation}%
These relations impose some additional constrains on the expansion coefficients $a_{n}$ and $b_{n}$.

Now let us consider the boundary condition at $r=R(t)$ for the functions $W(r,t)$,
$\Psi (r,t)$, and $\frac{\partial }{\partial r}\Psi (r,t)$. For the
function $W(r,t)$ this boundary condition yields $J_{0}(\kappa _{n})=0$,
i.e. the quantities $\kappa _{n}$ (with $n=1,2,...$) must be the positiv
zeros of the Bessel function $J_{0}(z)$. Later on we will assume that the
zeros $\kappa _{n}$ are arranged in ascending order of magnitude. The same
boundary condition for the quantity $\Psi (r,t)$ determines the unknown
function $C(t)$ in Eq.~\eqref{eq:10},
\begin{equation}
C(t)=-\frac{R(t)}{R_{0}^{2}}\sum_{n=1}^{\infty
}a_{n}A_{n}T_{n}(t)J_{1}(\lambda _{n}) .
\label{eq:21}
\end{equation}%
Finally, the boundary condition at $r=R(t)$ for $\frac{\partial }{\partial r} \Psi (r,t)$ yields another
relation for the function $C(t)$ which should be consistent with Eq.~\eqref{eq:21}. This is only possible
if $J_{0}(\lambda _{n})=0$, i.e. $\lambda _{n}=\kappa _{n}$ are also the zeros of the Bessel function.

Let us now turn to the determination of the expansion coefficients $a_{n}$ and $b_{n}$ using the constrains
in Eq.~\eqref{eq:20}. Inserting Eq.~\eqref{eq:14} into Eq.~\eqref{eq:20} and using the summation formulas~\eqref{eq:ap8}
and \eqref{eq:ap4} one can easily prove that the equations in Eq.~\eqref{eq:20} are satisfied if $a_{n}A_{n}=-4/[\lambda
_{n}^{2}J_{1}(\lambda _{n})]$ and $b_{n}B_{n}=-2/[\lambda _{n}J_{1}(\lambda _{n})]$. Then taking into account
the summation formulas \eqref{eq:ap9} and \eqref{eq:ap4} the initial conditions $C(0)=1$ and $W(r,0)=0$ (at
$r\leqslant R_{0}$) for the functions $C(t)$ and $W(r,t)$ imply that $T_{n}(0)=R_{0}$ and $U_{n}(0)=R_{0}^{2}$,
respectively. Moreover, having in mind the relations \eqref{eq:ap8} and \eqref{eq:ap9} the initial conditions
for the quantity $\Psi (r,0)$ inside ($r\leqslant R_{0}$) and outside ($r\geqslant R_{0}$) the plasma cylinder
(see Eq.~\eqref{eq:9}) are then satisfied automatically.

Therefore, the complete solution of Eqs.~\eqref{eq:4}-\eqref{eq:7} subject to
the initial and boundary conditions inside ($r\leqslant R(t)$) and outside ($%
r\geqslant R(t)$) the plasma cylinder, respectively, may be represented as
\begin{eqnarray}
&&\Psi (r,t)=r-4\sum_{n=1}^{\infty }T_{n}(t)\frac{J_{1}(\lambda _{n}\xi )}{%
\lambda _{n}^{2}J_{1}(\lambda _{n})},  \label{eq:22} \\
&&W(r,t)=r\left[ 1-\frac{2}{R^{2}(t)}\sum_{n=1}^{\infty }U_{n}(t)\frac{%
J_{0}(\lambda _{n}\xi )}{\lambda _{n}J_{1}(\lambda _{n})}\right] ,
\label{eq:23}
\end{eqnarray}%
and $W(r,t)=r$,
\begin{equation}
\Psi (r,t)=r-\frac{4R(t)}{r}\mathcal{T}(t)
\label{eq:24}
\end{equation}%
with
\begin{equation}
\mathcal{T}(t)=\sum_{n=1}^{\infty }\frac{1}{\lambda _{n}^{2}}T_{n}(t),  \quad
\mathcal{U}(t)=\sum_{n=1}^{\infty }\frac{1}{\lambda _{n}^{2}}U_{n}(t) .
\label{eq:24a}
\end{equation}%
Note that $\Psi (r,t)=W(r,t)=r$ at $r\to \infty $ as expected by the boundary conditions at the infinity.

The $\varphi $-component of the vector potential is determined by the first relation in Eq.~\eqref{eq:1}.
The straightforward integrations in Eq.~\eqref{eq:23} with respect to the radial coordinate $r$ result in \cite{gra80}
\begin{eqnarray}
&&A_{\varphi }(r,t)=\frac{H_{0\parallel }}{2}\left[ r-\frac{4}{R(t)}%
\sum_{n=1}^{\infty }U_{n}(t)\frac{J_{1}(\lambda _{n}\xi )}{\lambda
_{n}^{2}J_{1}(\lambda _{n})}\right] ,  \label{eq:25} \\
&&A_{\varphi }(r,t)=\frac{H_{0\parallel }}{2}\left[ r-\frac{G(t)}{r}\right]
\label{eq:26}
\end{eqnarray}%
inside ($r\leqslant R(t)$) and outside ($r\geqslant R(t)$) the plasma
cylinder, respectively. Here $G(t)$ is an arbitrary function of time to be
determined by the boundary condition at $r=R(t)$. From this condition one
obtains $G(t)=4\mathcal{U}(t)$. Equations \eqref{eq:22}-\eqref{eq:26}
represent the complete solution of the problem and determine
the structure of the electromagnetic fields both inside and outside the
expanding plasma cylinder. Expressions for the components of the
electromagnetic fields $\mathbf{E}$ and $\mathbf{H}$ may now be obtained by
use of Eqs.~\eqref{eq:2} and \eqref{eq:3}. These components are, for $r\leqslant R(t)$,
\begin{eqnarray}
&&H_{r}=H_{0\perp }\cos \varphi \left[ 1-\frac{4}{r}\sum_{n=1}^{\infty
}T_{n}(t)\frac{J_{1}(\lambda _{n}\xi )}{\lambda _{n}^{2}J_{1}(\lambda _{n})}%
\right] ,  \label{eq:28} \\
&&H_{\varphi }=-H_{0\perp }\sin \varphi \left[ 1-\frac{4}{R(t)}%
\sum_{n=1}^{\infty }T_{n}(t)\frac{J_{1}^{\prime }(\lambda _{n}\xi )}{\lambda
_{n}J_{1}(\lambda _{n})}\right] ,  \label{eq:29} \\
&&H_{z}=H_{0\parallel }\left[ 1-\frac{2}{R^{2} (t)}\sum_{n=1}^{\infty
}U_{n}(t)\frac{J_{0}(\lambda _{n}\xi )}{\lambda _{n}J_{1}(\lambda _{n})}%
\right] ,  \label{eq:30}
\end{eqnarray}
\begin{eqnarray}
&&E_{\varphi }=\frac{2H_{0\parallel }}{cR(t)}\sum_{n=1}^{\infty }\frac{1}{%
\lambda _{n}^{2}J_{1}(\lambda _{n})} \label{eq:31} \\
&&\times \left[ \dot{U}_{n}(t)J_{1}(\lambda
_{n}\xi )-\frac{\dot{R}(t)}{R(t)}U_{n}(t)\lambda _{n}\xi J_{0}(\lambda
_{n}\xi )\right] ,  \nonumber \\
&&E_{z}=\frac{4H_{0\perp }}{c}\sin \varphi \sum_{n=1}^{\infty }\frac{1}{%
\lambda _{n}^{2}J_{1}(\lambda _{n})} \label{eq:32} \\
&&\times \left[ \dot{T}_{n}(t)J_{1}(\lambda
_{n}\xi )-\frac{\dot{R}(t)}{R(t)}T_{n}(t)\lambda _{n}\xi J_{1}^{\prime
}(\lambda _{n}\xi )\right] ,  \nonumber
\end{eqnarray}%
and, for $r\geqslant R(t)$, $H_{z}=H_{0\parallel }$,
\begin{eqnarray}
&&H_{r}=H_{0\perp }\cos \varphi \left[ 1-\frac{4R(t)}{r^{2}}\mathcal{T}(t)%
\right] ,  \label{eq:33} \\
&&H_{\varphi }=-H_{0\perp }\sin \varphi \left[ 1+\frac{4R(t)}{r^{2}}\mathcal{%
T}(t)\right] ,  \label{eq:34}
\end{eqnarray}
\begin{eqnarray}
&&E_{\varphi }=\frac{H_{0\parallel }}{cr}\left[ R(t) \dot{R}(t) -\frac{2D(t)}{%
R^{2}(t)}\sum_{n=1}^{\infty }U_{n}(t)\right] ,  \label{eq:35} \\
&&E_{z}=\frac{H_{0\perp }}{cr}\sin \varphi \left\{ \dot{R}(t) [R(t) +4%
\mathcal{T}(t)] -\frac{4D(t)}{R(t)}\sum_{n=1}^{\infty
}T_{n}(t)\right\}.  \label{eq:36}
\end{eqnarray}%
In the latter expressions for the components of the electric field the time-derivatives $\dot{U}_{n}(t)$
and $\dot{T}_{n}(t)$ have been excluded by means of Eqs.~\eqref{eq:15} and \eqref{eq:16}. Also in
Eqs.~\eqref{eq:28}-\eqref{eq:36} the prime indicates the derivative of the Bessel function with
respect to the argument.

The induced current density $\mathbf{j}$ is defined only inside the plasma
cylinder (i.e., for $r\leqslant R(t)$) by the relation \eqref{eq:ohm} (or
using the Maxwell equation $\mathbf{j}=\frac{c}{4\pi }\boldsymbol{\nabla }%
\times \mathbf{H}$) and has the following components: $j_{r}=0$, and
\begin{eqnarray}
&&j_{\varphi }=-\frac{cH_{0\parallel }}{2\pi R^{3}(t)}\sum_{n=1}^{\infty
}U_{n}(t)\frac{J_{1}(\lambda _{n}\xi )}{J_{1}(\lambda _{n})},  \label{eq:37}
\\
&&j_{z}=-\frac{cH_{0\perp }}{\pi R^{2}(t)}\sin \varphi \sum_{n=1}^{\infty
}T_{n}(t)\frac{J_{1}(\lambda _{n}\xi )}{J_{1}(\lambda _{n})}.  \label{eq:38}
\end{eqnarray}

Until now we have considered the initial value problem (ii) assuming that initially the plasma conductivity
is so high that the magnetic field is completely excluded from the initial volume of a plasma. The extension
of the obtained solution to the case of the initial value problem (i) (with
highly resistive plasma at $t=0$) is straightforward. From the consideration
above it follows that the solution of the boundary and initial value problem
(i) is again determined by Eqs.~\eqref{eq:22}-\eqref{eq:38}, where, however,
the initial conditions $T_{n}(0)=R_{0}$ and $U_{n}(0)=R_{0}^{2}$ for the
functions $T_{n}(t)$ and $U_{n}(t)$ (see Eqs.~\eqref{eq:18} and \eqref{eq:19})
should be replaced by the zero initial conditions, $%
T_{n}(0)=U_{n}(0)=0$. Comparison of the complete solutions obtained by the
initial value problems (i) and (ii) shows that the electromagnetic fields
and the induced current for the two distinct cases differ only in the terms
containing $T_{n}(0)$ and $U_{n}(0)$ in Eqs.~\eqref{eq:18} and \eqref{eq:19}.
These terms force the matching of the solution to the initial
condition of the magnetic field completely excluded initially from the
plasma volume $r\leqslant R_{0}$. Because of their exponential dependence on
$\vartheta (t)$ (and hence on the time $t$) these terms become negligible
compared to the other terms in the electromagnetic fields in the time
required for $\vartheta (t)$ to become of the order $\vartheta (t)\sim 1$.
This time interval is the characteristic time for diffusion of the magnetic
field into a stationary plasma of radius $R$ and conductivity $\sigma $,
i.e., $\tau _{\mathrm{D}}\simeq R^{2}/D=4\pi \sigma R^{2}/c^{2}$. Thus the
initial conditions for the initially perfectly conducting plasma are
forgotten by the plasma at $t\gtrsim \tau _{\mathrm{D}}$.

In the context of the two distinct initial value problems (i) and (ii) it
should be also emphasized that the initial value of the plasma conductivity $%
\sigma (0)$ should be consistent with the chosen physical model. Indeed, the
cases (i) and (ii) imply vanishing ($\sigma (0)\to 0$) and very
large ($\sigma (0)\to \infty $) initial conductivities of the
expanding plasma, respectively. As a demonstration of the importance of the
initial value $\sigma (0)$ consider, for instance, Eqs.~\eqref{eq:31}, %
\eqref{eq:32} and \eqref{eq:35}, \eqref{eq:36}\ for the generated electric
field. Using Eqs.~\eqref{eq:15} and \eqref{eq:16} as well as the summation
formulas of Appendix~\ref{sec:app1} it is straightforward to show that at
$t=0$ the electric field inside (Eqs.~\eqref{eq:31} and \eqref{eq:32}) the
plasma cylinder is given by
\begin{eqnarray}
&&E_{\varphi }(0)=r\frac{H_{0\parallel }}{c}\frac{\dot{R}_{0}}{R_{0}}\left(
1-\frac{u_{0}}{R_{0}^{2}}\right) ,  \label{eq:39} \\
&&E_{z}(0)=r\frac{H_{0\perp }}{c}\frac{\dot{R}_{0}}{R_{0}}\sin \varphi
\left( 1-\frac{t_{0}}{R_{0}}\right) ,  \label{eq:40}
\end{eqnarray}%
where $\dot{R}_{0}=\dot{R}(0)$. Thus, the initial electric field in the plasma volume vanishes or
is finite in the cases (ii) (with $u_{0}=R_{0}^{2}$, $t_{0}=R_{0}$) and (i)
(with $u_{0}=t_{0}=0$), respectively. The initial electric field in a vacuum
is determined by Eqs.~\eqref{eq:35} and \eqref{eq:36} at $t=0$.
It is seen that at $t\to 0$ the last terms in these
expressions proportional to the diffusion coefficient $D(t)$ vanish for the
initial condition (i) while diverging as $\sim \lbrack \vartheta (t)]^{-1/2}$
in the case (ii). In the latter case assuming, however, a perfectly
conducting (in fact infinitely conducting) plasma at $t=0$ one must consider
the limit $D(0)\to 0$ in the last terms of Eqs.~\eqref{eq:35} and %
\eqref{eq:36} which vanish eventually at $t\to 0$. Finally, we note
that at $t\to 0$ the nonvanishing terms in Eqs.~\eqref{eq:35}, %
\eqref{eq:36} and \eqref{eq:39}, \eqref{eq:40} are proportional to the
initial expansion velocity $\dot{R}_{0}$ of the plasma. Therefore, it is not
surprising that at $\dot{R}_{0}\neq 0$ the plasma expansion builds up
instantly an initial electric field although the induced magnetic field is
zero.

At the end of this section consider briefly the nondiffusive limit of the
obtained solutions, Eqs.~\eqref{eq:28}-\eqref{eq:38}, when the diffusion
coefficient vanishes, $D\to 0$. This limit can be obtained using at $%
D\to 0$ the expressions~\eqref{eq:17}-\eqref{eq:19} which yield $%
T_{n}(t)=R(t)$ and $U_{n}(t)=R^{2}(t)$. Therefore, having in mind the
summation formula~\eqref{eq:ap9} from Eq.~\eqref{eq:24a} one finds $\mathcal{%
T}(t)=R(t)/4$ and $\mathcal{U}(t)=R^{2}(t)/4$. Using these results and the
summation formulas derived in Appendix~\ref{sec:app1} it is straightforward
to show that $j_{\varphi }=j_{z}=0$ and the electromagnetic fields are, for
$r\leqslant R(t)$, $\mathbf{H}(\mathbf{r},t)=\mathbf{E}(\mathbf{r},t)=0$,
and, for $r\geqslant R(t)$, $H_{z}=H_{0\parallel }$, $H_{r}=H_{0\perp }\cos
\varphi (1-R^{2}/r^{2})$, $H_{\varphi }=-H_{0\perp }\sin \varphi
(1+R^{2}/r^{2})$, $E_{\varphi }=\beta H_{0\parallel }(R/r)$, $E_{z}=2\beta
H_{0\perp }(R/r)\sin \varphi $, where $\beta =\dot{R}/c$. These expressions
have been derived previously in Ref.~\cite{ner10} for the cylindrical plasma
expansion neglecting the diffusion of the magnetic field.

\section{Energy balance}
\label{sec:4}

Previously significant attention has been paid \cite{osi03,ner09,rai63,dit00,ner06,ner12,ner10}
to the question of what fraction of energy is emitted and lost in the form of electromagnetic
pulse propagating outward of the expanding plasma. In this section we consider the energy balance
during the plasma cylinder expansion in the presence of the homogeneous magnetic field.

Our starting point is the energy balance equation
\begin{equation}
\mathbf{\nabla }\cdot \mathbf{S}=-\mathbf{j}\cdot \mathbf{E}-\frac{\partial
}{\partial t}\frac{H^{2}}{8\pi } ,
\label{eq:41}
\end{equation}%
where $\mathbf{S}=\frac{c}{4\pi }[\mathbf{E}\times \mathbf{H}]$ is the
Poynting vector and $\mathbf{j}$ is the induced current. Note that the
density of the electric field energy has been neglected in Eq~\eqref{eq:41}
since $\dot{R}\ll c$ and $E\ll H$. The energy emitted to infinity is
measured as a Poynting vector integrated over time and over the lateral
surface $S_{c}$ of the cylinder with radius $r_{c}$, length $l_{c}$ and the
volume $\Omega _{c}$ (control cylinder) enclosing the plasma cylinder ($%
r_{c}>R(t)$). Integrating over time and over the volume $\Omega _{c}$
Eq.~\eqref{eq:41} can be represented as
\begin{equation}
W_{\mathrm{S}}(t)=W_{\mathrm{J}}(t)+\Delta W_{\mathrm{M}}(t) ,
\label{eq:42}
\end{equation}%
where
\begin{eqnarray}
&&W_{\mathrm{S}}(t)=r_{c}\int_{0}^{t}dt^{\prime }\int_{0}^{2\pi }S_{r}d\varphi ,  \nonumber \\
&&W_{\mathrm{J}}(t)=-\frac{1}{l_{c}}\int_{0}^{t}dt^{\prime }\int_{\Omega _{c}}\mathbf{j}\cdot \mathbf{E}d\mathbf{r} . \label{eq:43}
\end{eqnarray}%
Here $S_{r}$ is the radial component of the Poynting vector. Note that the total flux of the
energy over the bases of the control cylinder determined by $S_{z}$ vanishes due to the symmetry reason.
$W_{\mathrm{M}} (t)$ and $\Delta W_{\mathrm{M}}(t)=W_{\mathrm{M}}(0)-W_{\mathrm{M}}(t)$ are
the total magnetic energy and its change (with minus sign) per unit length
in a volume $\Omega _{c}$, respectively. $W_{\mathrm{J}}(t)$ is the energy
(per unit length) transferred from plasma cylinder to the magnetic field and
is the mechanical work with minus sign performed by the plasma on the
external magnetic pressure. At $t=0$ the magnetic fields are given by
$\mathbf{H} =\mathbf{H}_{0}$ in the model (i) and by Eq.~\eqref{eq:9}
in the model (ii). Hence in (i) $W_{\mathrm{M}}(0)$ is the
total magnetic energy per unit length in a volume $\Omega _{c}$ and is given
by $W_{\mathrm{M}}(0)\equiv Q_{c}=\pi r_{c}^{2}(H_{0}^{2}/8\pi )$ while in
(ii) $W_{\mathrm{M}}(0)$ is the total magnetic energy in a volume $\Omega
_{c}^{\prime }$ of the control cylinder excluding the volume of the plasma
cylinder (we take into account that $\mathbf{H} =0$ at $t=0$
in a plasma cylinder in the model (ii))\ and $W_{\mathrm{M}%
}(0)=Q_{c}(1-\kappa ^{-2}\cos ^{2}\theta -\kappa ^{-4}\sin ^{2}\theta )$.
Here $\kappa =r_{c}/R_{0}$ and $\theta $ is the angle between $\mathbf{H}%
_{0}$ and the symmetry axis of the plasma cylinder ($z$-axis). Then the
change of the magnetic energy $\Delta W_{\mathrm{M}}(t)$ in a volume $\Omega_{c}$
can be evaluated as
\begin{equation}
\Delta W_{\mathrm{M}}(t)=W_{\mathrm{M}}(0)-\frac{1}{l_{c}}\int_{\Omega _{c}}%
\frac{H^{2}}{8\pi }d\mathbf{r} .
\label{eq:44}
\end{equation}%
Hence the total energy flux $W_{\mathrm{S}}(t)$ given by Eq.~\eqref{eq:43}
is calculated as a sum of the energy loss by plasma due to the external
magnetic pressure and the decrease of the magnetic energy in a control
volume $\Omega _{c}$. For expansion of a one-dimensional plasma slab and for
uniform external magnetic field $W_{\mathrm{S}}\simeq 2W_{\mathrm{J}}\simeq
2\Delta W_{\mathrm{M}}$, i.e., approximately the half of the outgoing energy
is gained from the plasma, while the other half is gained from the magnetic
energy \cite{dit00}. In the case of expansion of highly conducting spherical
plasma with radius $R$ in the uniform magnetic field $\mathbf{H}_{0}$ the
outgoing energy $W_{\mathrm{S}} $ is distributed between $W_{\mathrm{J}}$
and $\Delta W_{\mathrm{M}}$ according to $W_{\mathrm{J}}=1.5Q_{0}$ and $%
\Delta W_{\mathrm{M}}=0.5Q_{0}$ with $W_{\mathrm{S}}=2Q_{0}$, where $%
Q_{0}=H_{0}^{2}R^{3}/6$ is the magnetic energy escaped from the spherical
plasma volume \cite{rai63}. Therefore, in this case the released magnetic
energy is mainly gained from the plasma.

Consider now each energy component $W_{\mathrm{S}}(t)$, $W_{\mathrm{J}}(t)$
and $\Delta W_{\mathrm{M}}(t)$ separately. $W_{\mathrm{S}}(t)$ is calculated
from Eq.~\eqref{eq:43} using the expressions for the electromagnetic fields
generated outside the plasma ($r\geqslant R(t)$). Recalling that $H_{z}=H_{0\parallel }$
at $r\geqslant R(t)$ from Eqs.~\eqref{eq:34}-\eqref{eq:36} and \eqref{eq:43} we obtain
\begin{eqnarray}
&&W_{\mathrm{S}}(t)=H_{0\parallel }^{2}\left[ \mathcal{U}(t)-\frac{u_{0}}{4}%
\right] +H_{0\perp }^{2}\bigg[ R(t)\mathcal{T}(t)    \nonumber  \\
&& +\frac{2R^{2}(t)}{R_{0}^{2}\kappa
^{2}}\mathcal{T}^{2}(t)-R_{0}\frac{t_{0}}{4}-\frac{t_{0}^{2}}{8\kappa ^{2}} \bigg] .  \label{eq:45}
\end{eqnarray}%
Here $\mathcal{T}(t)$ and $\mathcal{U}(t)$ are given by Eq.~\eqref{eq:24a}.

Next, we evaluate the energy loss $W_{\mathrm{J}}(t)$ by the plasma which is
determined by the induced current density, $\mathbf{j}$. This current has
two azimuthal and axial components and has been determined in Sec.~\ref{sec:3},
see Eqs.~\eqref{eq:37} and \eqref{eq:38}. Since the current is
localized only within a plasma volume $\Omega _{\mathrm{R}}$, in Eq.~\eqref{eq:43}
the volume $\Omega _{c}$ can be replaced by the plasma volume $\Omega _{\mathrm{R}}$.
The total energy loss by the plasma cylinder is calculated as
\begin{equation}
W_{\mathrm{J}}(t)=\frac{H_{0\parallel }^{2}}{2}\left[ \frac{\mathfrak{U}(t)}{%
R^{2}(t)}-\frac{u_{0}^{2}}{4R_{0}^{2}}\right] +H_{0\perp }^{2}\left[
\mathfrak{T}(t)-\frac{t_{0}^{2}}{4}\right] ,
\label{eq:46}
\end{equation}%
where
\begin{equation}
\mathfrak{T}(t)=\sum_{n=1}^{\infty }\frac{1}{\lambda _{n}^{2}}%
T_{n}^{2}(t),\quad \mathfrak{U}(t)=\sum_{n=1}^{\infty }\frac{1}{\lambda
_{n}^{2}}U_{n}^{2}(t) .
\label{eq:47}
\end{equation}

\begin{figure*}[tbp]
\begin{center}
\includegraphics[width=55.0mm]{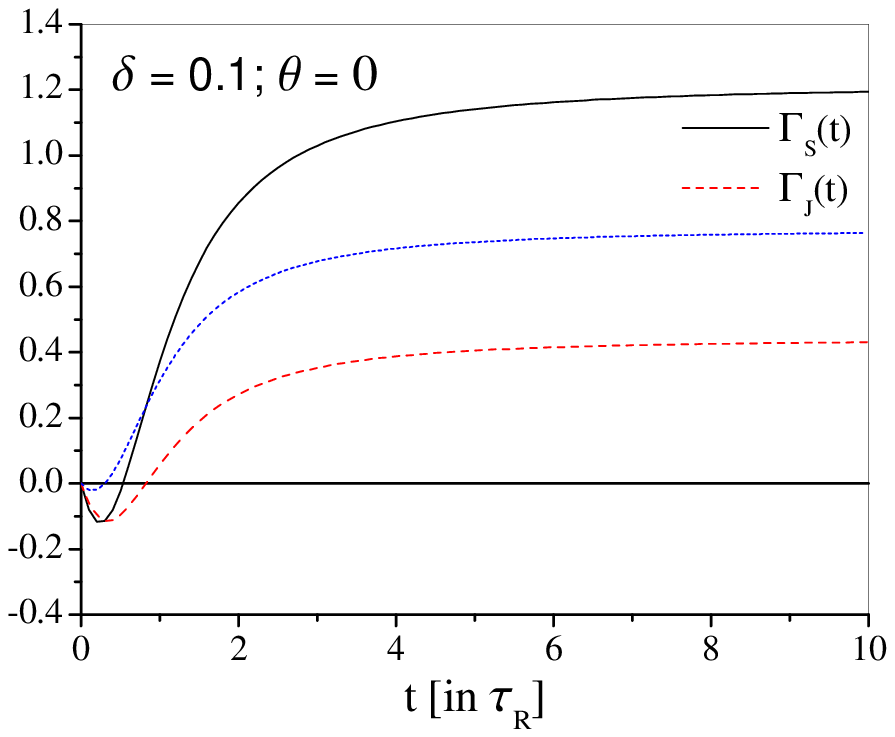} %
\includegraphics[width=55.0mm]{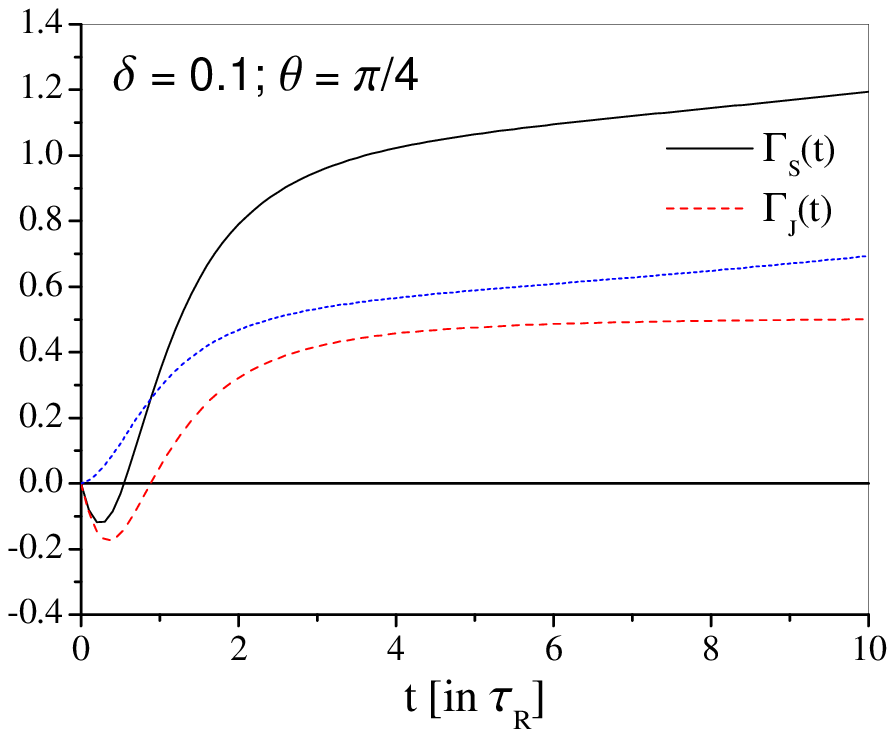} %
\includegraphics[width=55.0mm]{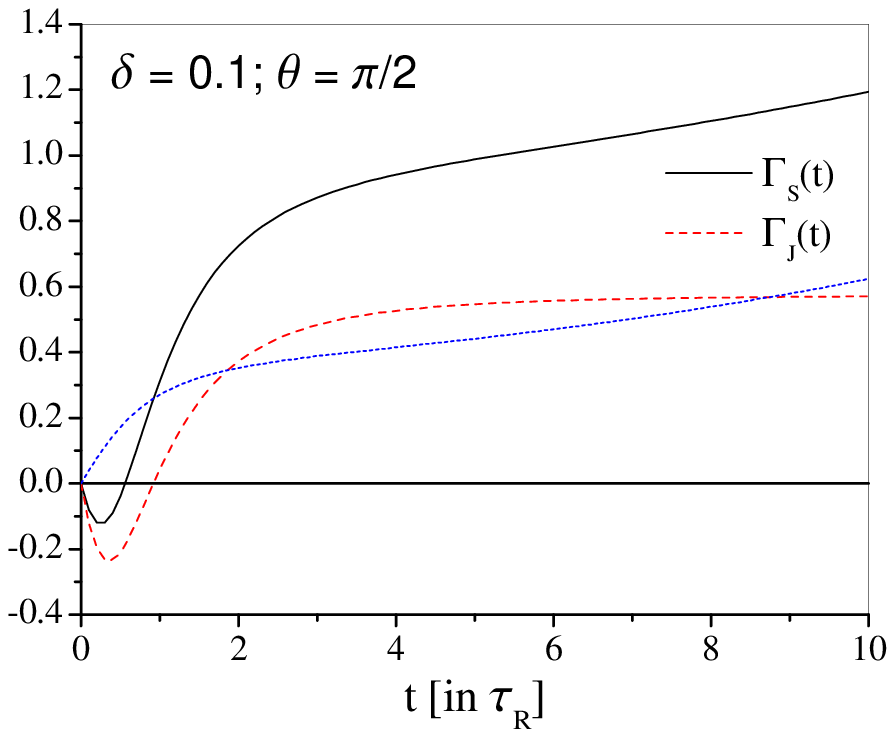}
\end{center}
\caption{(Color online) The ratios $\Gamma _{\mathrm{S}}(t)$ (solid lines) and $\Gamma _{\mathrm{J}}(t)$ (dashed lines)
for $\kappa =10$ and $\delta =0.1 $ as a function of $t$ (in units of $\tau_{\mathrm{R}} =R_{0}/v$) calculated
from expressions \eqref{eq:45} and \eqref{eq:46} with $\theta =0$ (left panel), $\theta =\pi /4$ (middle panel),
and $\theta =\pi /2$ (right panel). The dotted lines represent the quantity $\Gamma _{\mathrm{M}}(t) =\Gamma%
_{\mathrm{S}}(t) -\Gamma _{\mathrm{J}}(t)$.}
\label{fig:1}
\end{figure*}

The change of the magnetic energy in a control cylinder is calculated from Eq.~\eqref{eq:44}. For evaluation
of the magnetic energy inside and outside the plasma volume we use Eqs.~\eqref{eq:28}-\eqref{eq:30} and
\eqref{eq:33}, \eqref{eq:34} together with $H_{z}=H_{0\parallel }$, respectively. Thus the change of the total
magnetic energy in the control cylinder is represented as
\begin{eqnarray}
&&\Delta W_{\mathrm{M}}(t)=\frac{H_{0\parallel }^{2}}{2}\left[ 2\mathcal{U}%
(t)-\frac{\mathfrak{U}(t)}{R^{2}(t)}-\frac{u_{0}}{4}\left( 2-\frac{u_{0}}{%
R_{0}^{2}}\right) \right]  \nonumber \\
&&+H_{0\perp }^{2}\left\{ R(t)\mathcal{T}(t) -\mathfrak{T}(t) +2\mathcal{T}^{2}(t)
\frac{R^{2}(t)}{R_{0}^{2}\kappa ^{2}}
-\frac{t_{0}}{8}\left[ \frac{%
t_{0}}{\kappa ^{2}}+2\left( R_{0}-t_{0}\right) \right]\right\} . \label{eq:48}
\end{eqnarray}%
Comparing now Eqs.~\eqref{eq:45}, \eqref{eq:46} and \eqref{eq:48} we conclude that $\Delta W_{\mathrm{M}}(t)
+W_{\mathrm{J}}(t)=W_{\mathrm{S}}(t)$ as predicted by the energy balance equation~\eqref{eq:42}. Let us recall
that Eqs.~\eqref{eq:45}, \eqref{eq:46} and \eqref{eq:48} are valid for both initial conditions (i) and (ii)
choosing appropriate values for the quantities $t_{0} $ and $u_{0}$ (i.e., $t_{0}=u_{0}=0$ in (i) and $t_{0}=R_{0}$,
$u_{0}=R_{0}^{2}$ in (ii)).

The energy components $W_{\mathrm{S}}(t)$ and $\Delta W_{\mathrm{M}}(t)$ depend naturally
on the radius $r_{c}$ of the control cylinder while $W_{\mathrm{J}}(t)$ is determined only
by the volume of the plasma cylinder. At $r_{c} \to \infty$ Eqs.~\eqref{eq:45} and \eqref{eq:48}
are finite and in this case $W_{\mathrm{S}}(t)$ represents the electromagnetic energy emitted
to infinity. It should be noted that at $r\to \infty$ the induced magnetic and electric fields
behave as $H\sim r^{-2}$ and $E\sim r^{-1}$, respectively, see Eqs.~\eqref{eq:33}--\eqref{eq:36}.
Consequently, at $r_{c} \to \infty$ the induced magnetic field does not contribute to the energy
flux $W_{\mathrm{S}}(t)$ and the latter is determined by the electric field and the unperturbed
magnetic field $\mathbf{H}_{0}$.

\section{Numerical examples}
\label{sec:5}

In this section using theoretical findings of the preceding sections, we present the results of our model
calculations for the electromagnetic fields generated due to the radial expansion of the magnetized
cylindrical plasma into a vacuum. As an example we consider the radial expansion with $R^{2}(t)=R_{0}^{2}
+v^{2}t^{2}$, where $v$ is the expansion velocity at $t\to \infty $. Note that initially $\dot{R}_{0}=0$
in this case. The electrical conductivity is modeled by $\sigma (t) =\sigma _{0}(\tau /t)$,
where $\tau $ is some characteristic decay time of $\sigma (t)$ and $\sigma _{0}$
is a constant. Accordingly the diffusion coefficient is given by $D(t)=D_{0}(t/\tau )$ with
$D_{0}=c^{2}/4\pi \sigma _{0}$. Thus, initially the conductivity of the plasma is very high which corresponds
to the initial condition (ii). However, as plasma expands the plasma will be eventually cooled off and the
conductivity decreases with time. For the chosen model $\vartheta (t)=\delta \ln [\eta (t)]$ as it follows
from Eq.~\eqref{eq:17}. Here $\delta =D_{0}/v^{2}\tau $ is the dimensionless diffusion coefficient
while $\eta (t)=R(t)/R_{0}$ is the dimensionless plasma radius. In order to obtain a physical idea of the
length and time scales involved, let us consider briefly a numerical example. Taking, for instance, $R_{0} =1$~mm
and $v= 10^{6}$~cm/s, one obtains $\tau_{\mathrm{R}} =R_{0}/v =0.1$~$\mu$s. For a laser generated plasma with
a temperature $T\sim 10^{5}$~K, one can take $\sigma_{0} \simeq 5\times 10^{14}$~s$^{-1}$ and $D_{0} \simeq
2\times 10^{5}$~cm$^{2}$/s. Using these parameters the dimensionless diffusion coefficient $\delta =1$ implies
the characteristic decay time $\tau \simeq 0.18$~$\mu$s for the conductivity.

Equations~\eqref{eq:18} and \eqref{eq:19} can be evaluated in a closed form
\begin{eqnarray}
&&T_{n}(t)=\frac{t_{0}}{[\eta (t)]^{\lambda _{n}^{2}\delta }}+R_{0}\frac{%
\eta (t)-[\eta (t)]^{-\lambda _{n}^{2}\delta }}{\lambda _{n}^{2}\delta +1},
\label{eq:49} \\
&&U_{n}(t)=\frac{u_{0}}{[\eta (t)]^{\lambda _{n}^{2}\delta }}+R_{0}^{2}\frac{%
\eta ^{2}(t)-[\eta (t)]^{-\lambda _{n}^{2}\delta }}{\lambda _{n}^{2}\delta
/2+1}.  \label{eq:50}
\end{eqnarray}%
It is seen that the first terms of Eqs.~\eqref{eq:49} and \eqref{eq:50} containing the initial values $t_{0}$
and $u_{0}$ vanish exponentially at $t\to \infty $. The electromagnetic fields, the induced currents and the
components of the energy are determined by inserting Eqs.~\eqref{eq:49} and \eqref{eq:50} into
Eqs.~\eqref{eq:28}--\eqref{eq:38} and Eqs.~\eqref{eq:45}--\eqref{eq:48}, respectively.

\begin{figure*}[tbp]
\begin{center}
\includegraphics[width=55.0mm]{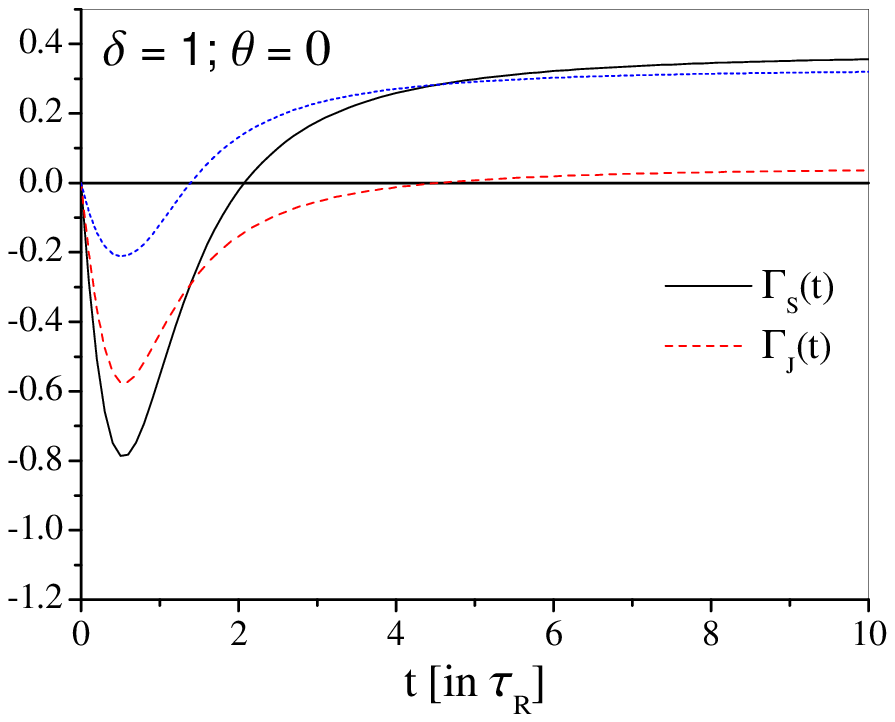} %
\includegraphics[width=55.0mm]{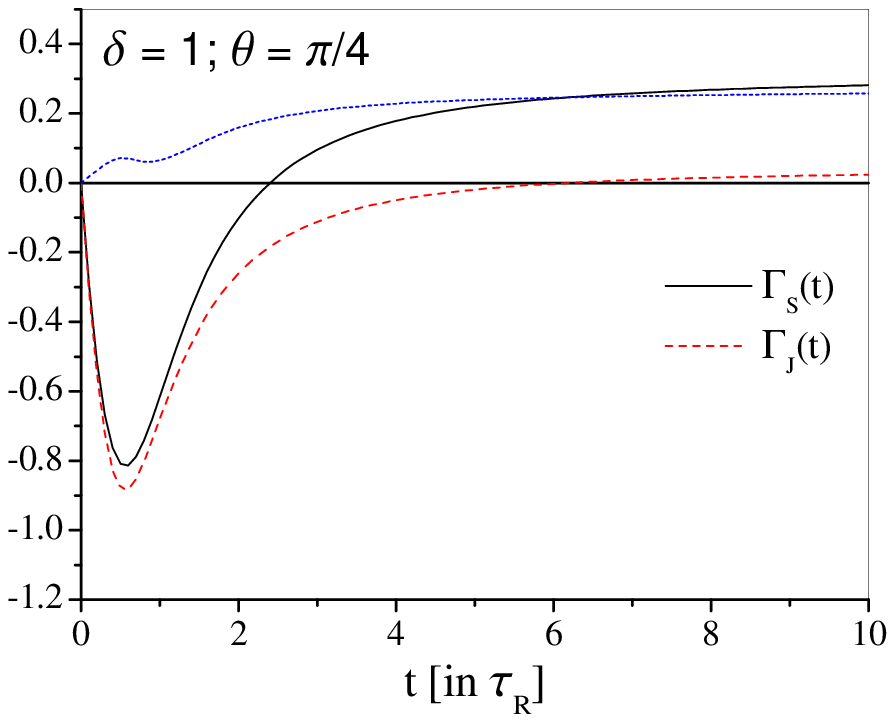} %
\includegraphics[width=55.0mm]{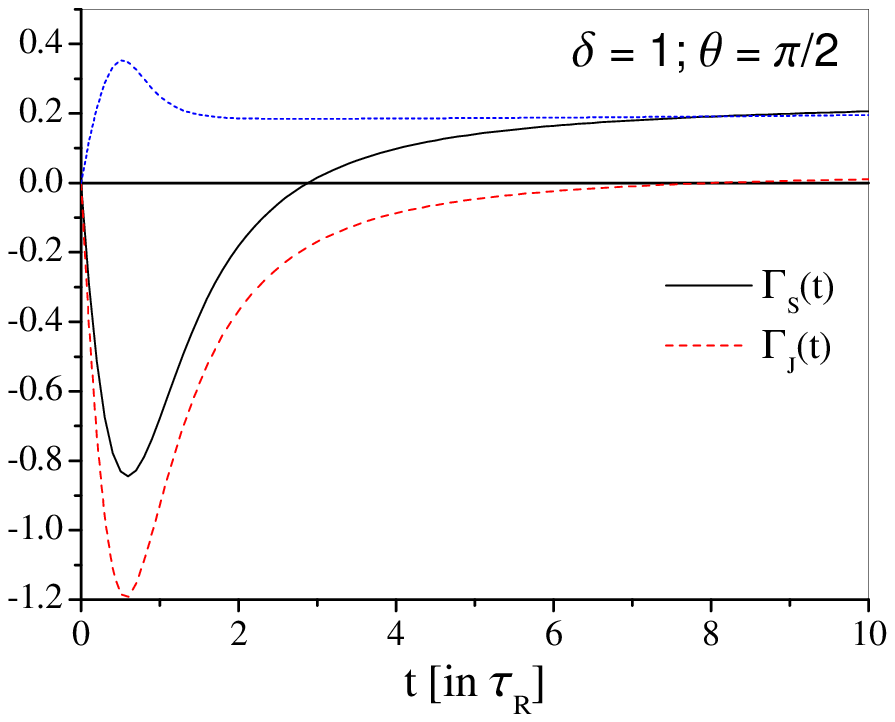}
\end{center}
\caption{(Color online) Same as in Fig.~\ref{fig:1} but for $\delta =1$.}
\label{fig:2}
\end{figure*}

\begin{figure*}[tbp]
\begin{center}
\includegraphics[width=55.0mm]{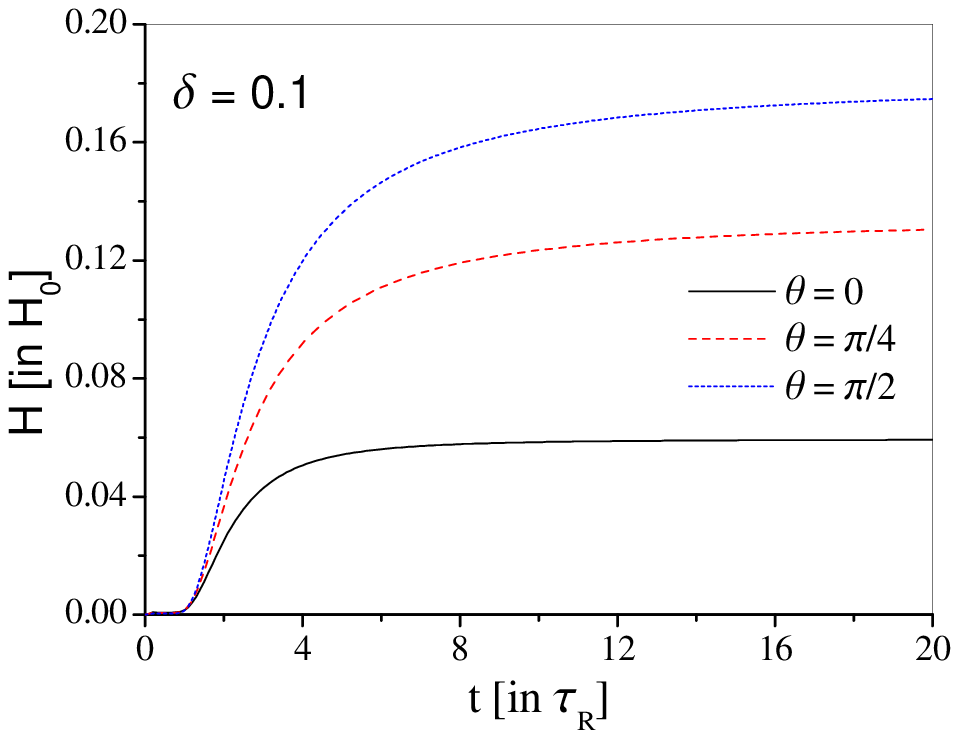} %
\includegraphics[width=55.0mm]{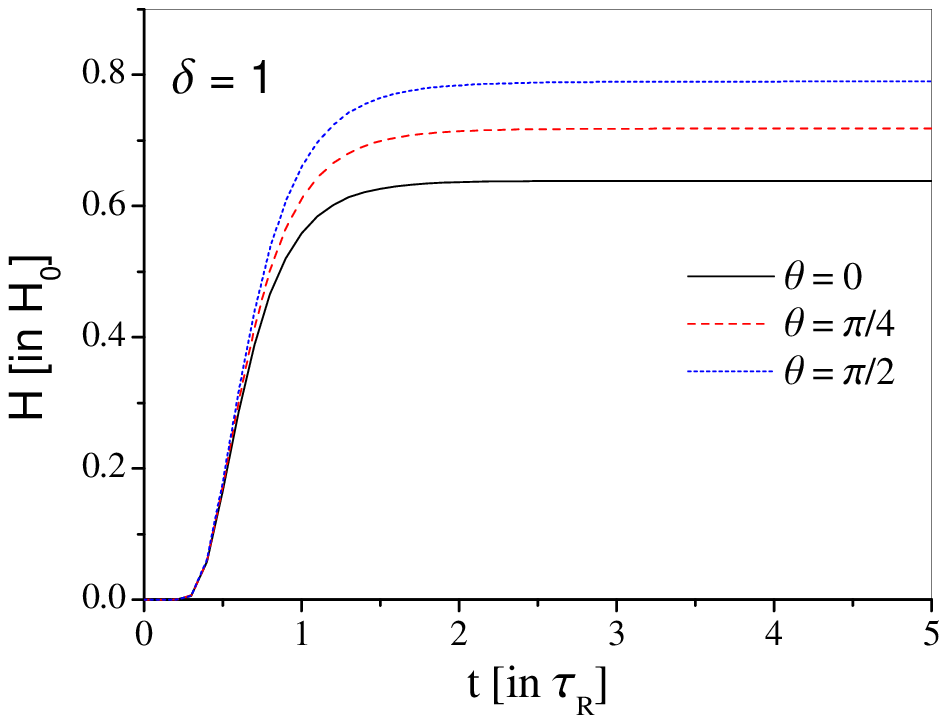} %
\includegraphics[width=55.0mm]{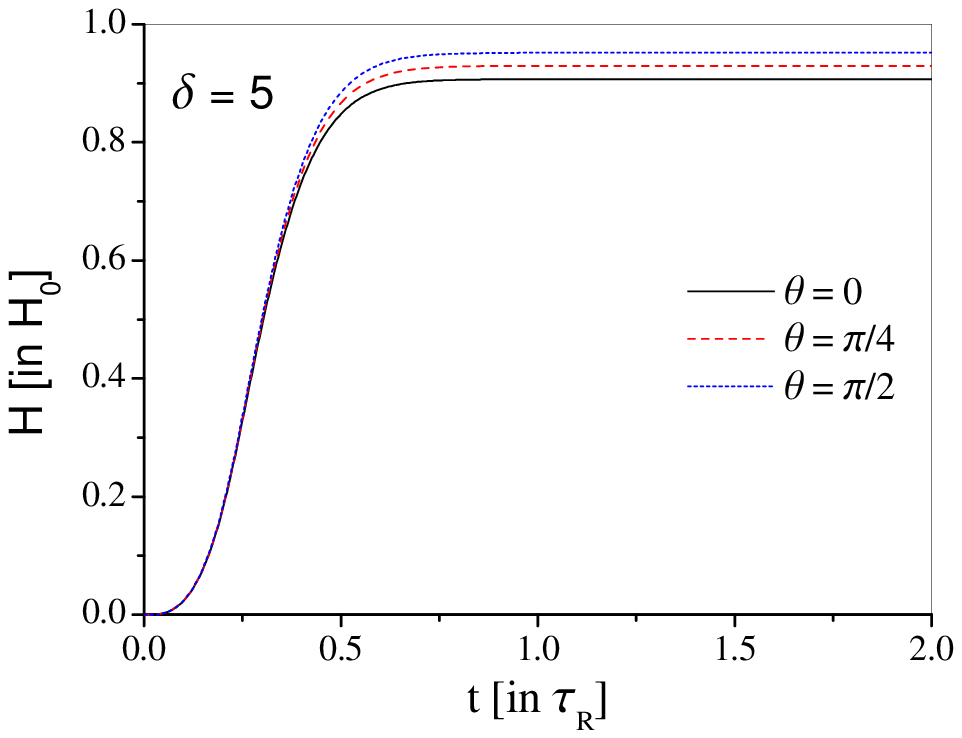}
\end{center}
\caption{(Color online) The strength of the magnetic field (in units of $H_{0}$) at $r=0$ as a function of time
$t$ (in units of $\tau_{\mathrm{R}}$) for $\delta =0.1$ (left panel), $\delta =1$ (middle panel), and $\delta =5$
(right panel). The solid, dashed and dotted lines correspond to $\theta =0$, $\theta =\pi /4$ and $\theta =\pi /2$,
respectively.}
\label{fig:3}
\end{figure*}

First, we consider the energy balance during the plasma expansion. As an example in Figs.~\ref{fig:1} and \ref{fig:2}
we show the results of our model calculations for the ratios $\Gamma _{\mathrm{S}}(t)=W_{\mathrm{S}}(t)/Q(t)$,
$\Gamma _{\mathrm{J}}(t)=W_{\mathrm{J}}(t)/Q(t)$ and $\Gamma _{\mathrm{M}}(t) = \Delta W_{\mathrm{M}}(t)/Q(t) =%
\Gamma _{\mathrm{S}}(t) -\Gamma _{\mathrm{J}}(t)$ as a function of time (scaled by $\tau_{\mathrm{R}} =R_{0}/v$) for
$\kappa =r_{c}/R_{0} =10$, and for various orientations of the magnetic field $\mathbf{H}_{0}$, $\theta =0$,
$\theta =\pi /4$, and $\theta =\pi /2$. Let us recall that Eqs.~\eqref{eq:45} and \eqref{eq:48} are valid at
$0\leqslant t/\tau_{\mathrm{R}}\leqslant (\kappa^{2} -1)^{1/2}$ (or $\eta (t) \leqslant \kappa$). Here $Q(t)=(\pi
R^{2}(t))(H_{0}^{2}/8\pi )$ is the total energy (per unit length) of the unperturbed magnetic field at time $t$ in the volume of the
plasma cylinder. For the dimensionless diffusion coefficient $\delta $ we have chosen two distinct values, $\delta =0.1$
(weak diffusion) and $\delta =1$ (strong diffusion). Unlike the cases with uniform magnetic field and highly conducting
spherical or one-dimensional plasmas mentioned in Sec.~\ref{sec:4} (see also Refs.~\cite{dit00,rai63}) there are no,
in general, simple relations between the energy components $W_{\mathrm{S}}(t)$, $W_{\mathrm{J}}(t)$ and $Q(t)$. Moreover,
in the present case with a finite conductivity of the plasma the ratio $\Gamma _{\mathrm{J}}(t)$ is not a constant
anymore (see Figs.~\ref{fig:1} and \ref{fig:2}) as predicted in Ref.~\cite{ner10} for highly conducting cylindrical
plasma. It has been shown in Ref.~\cite{ner10} that at $\sigma \to \infty $ the energy flux $W_{\mathrm{S}}(t)$ is
systematically positive and $\Gamma _{\mathrm{J}}(t)\sim 1$, i.e. the energy loss by a plasma is approximately equal
to the magnetic energy escaped from the highly conducting plasma volume (formation of a diamagnetic cavity). As
demonstrated in Figs.~\ref{fig:1} and \ref{fig:2} both energies $W_{\mathrm{J}}$ and $W_{\mathrm{S}}$ are negative
here at the initial stage (at $t\lesssim \tau_{\mathrm{R}}$) of the plasma expansion. That is, the flux of the
electromagnetic energy over the surface of the control cylinder is negative and the plasma gains energy from the
generated electromagnetic fields. It is noteworthy that the energy gained by the plasma is maximal for transverse
orientation of $\mathbf{H}_{0}$. The physical origin of these features is the diffusion of the magnetic field in an
expanding plasma. At $t=0$ the magnetic field is completely excluded from the plasma volume. However, as soon as the
plasma expands and the electrical conductivity is reduced the magnetic field is effectively diffused into the volume
of the expanding plasma giving rise to the negative energy flux over the control surface. Clearly this process is
intensified by increasing the diffusion coefficient as shown in Fig.~\ref{fig:2}. At the later time, $t\gtrsim
\tau_{\mathrm{R}}$, with essentially reduced $\sigma (t)$ the magnetic field almost freely fills the whole volume of
the plasma and both energies $W_{\mathrm{J}}$ and $W_{\mathrm{S}}$ become positive in this case. That is, the expanding
plasma loses its energy due to the transfer of the energy to the electromagnetic fields. At $t\gtrsim \tau_{\mathrm{R}}$
the quantities $\Gamma _{\mathrm{J}}(t)$ and $\Gamma _{\mathrm{S}}(t)$ are determined explicitly by
\begin{eqnarray}
&&\Gamma _{\mathrm{J}}(t)\simeq \Xi _{2}(\delta /2)\cos ^{2}\theta +2\Xi _{2}(\delta
)\sin ^{2}\theta ,  \label{eq:51} \\
&&\Gamma _{\mathrm{S}}(t)\simeq 2\Xi _{1} (\delta /2) \cos ^{2}\theta +\Xi _{3}(\delta )
\sin ^{2}\theta ,  \label{eq:52}
\end{eqnarray}
where $\Xi _{3}(\delta ) = 2\Xi _{1}(\delta )+(\eta ^{2}/\kappa ^{2})\Xi_{1}^{2}(\delta )$ and the functions $\Xi _{1}(z)$
and $\Xi _{2}(z)$ are given by Eqs.~\eqref{eq:ap13} and \eqref{eq:ap14}, respectively.
Note that the energies $W_{\mathrm{J}}(t)$ and $W_{\mathrm{S}}(t)$ are reduced by the magnetic field diffusion which at
$\delta >1$ results in $W_{\mathrm{J}}(t) \ll W_{\mathrm{S}}(t)$.

\begin{figure*}[tbp]
\begin{center}
\includegraphics[width=55.0mm]{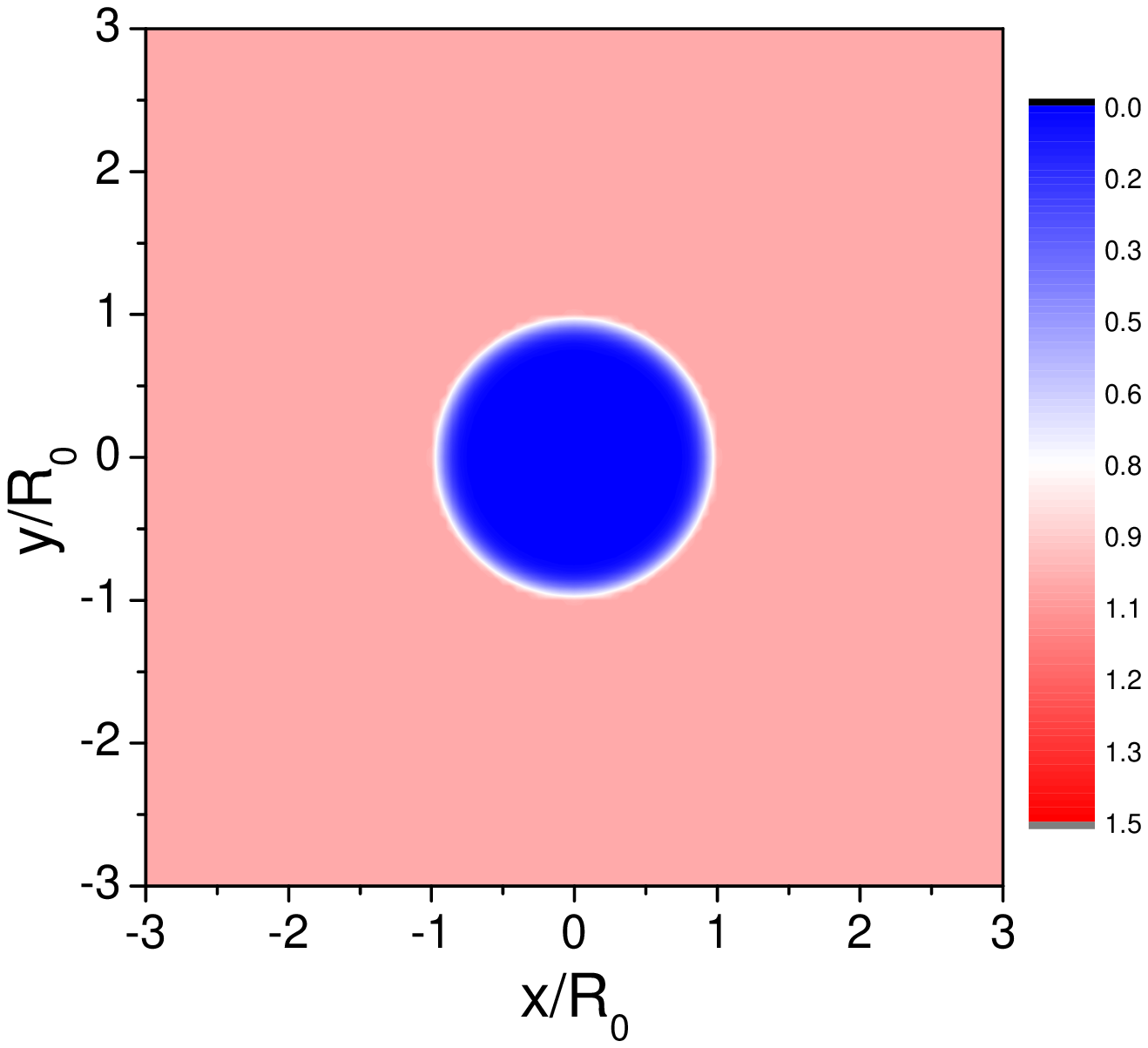} %
\includegraphics[width=55.0mm]{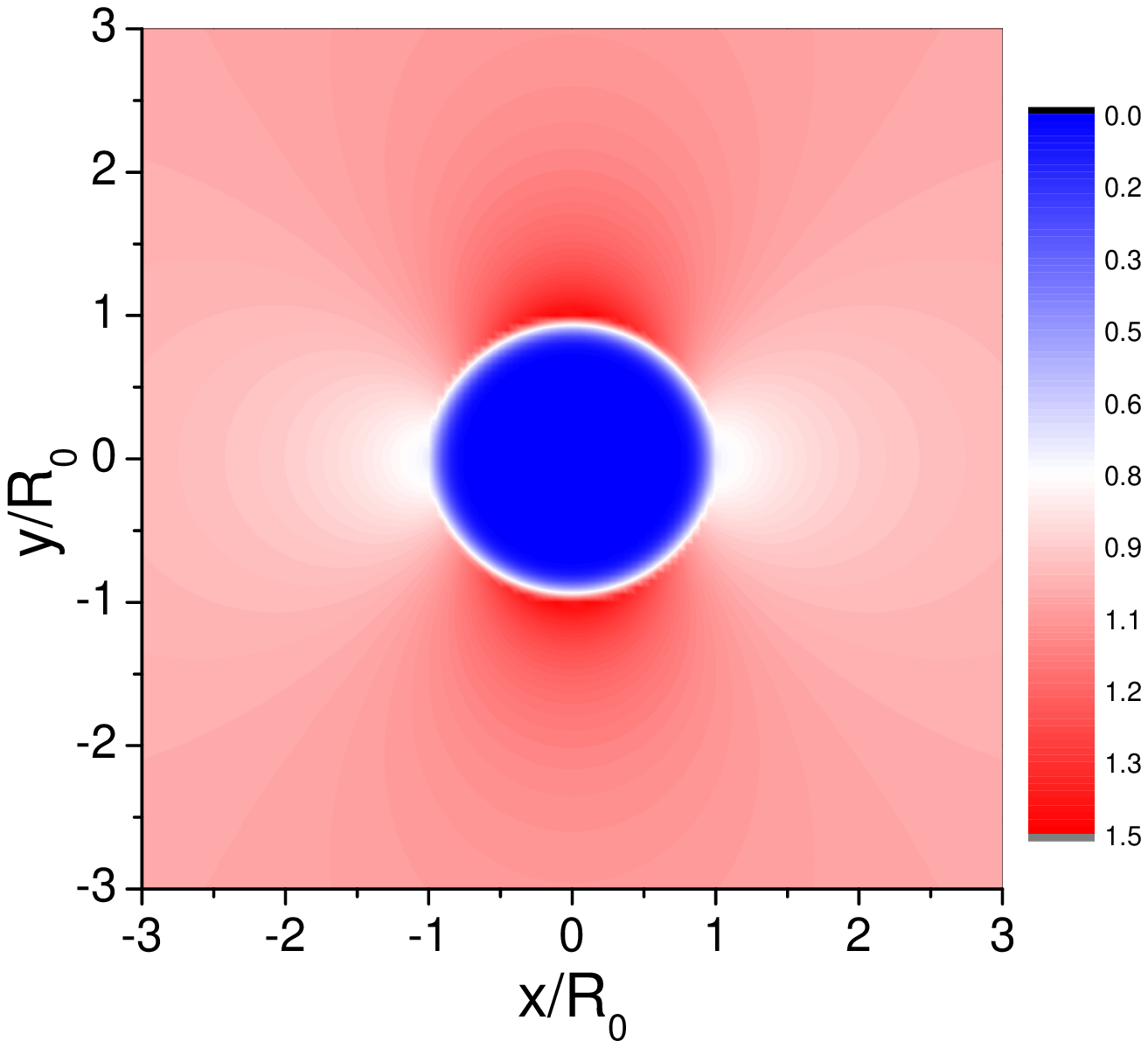} %
\includegraphics[width=55.0mm]{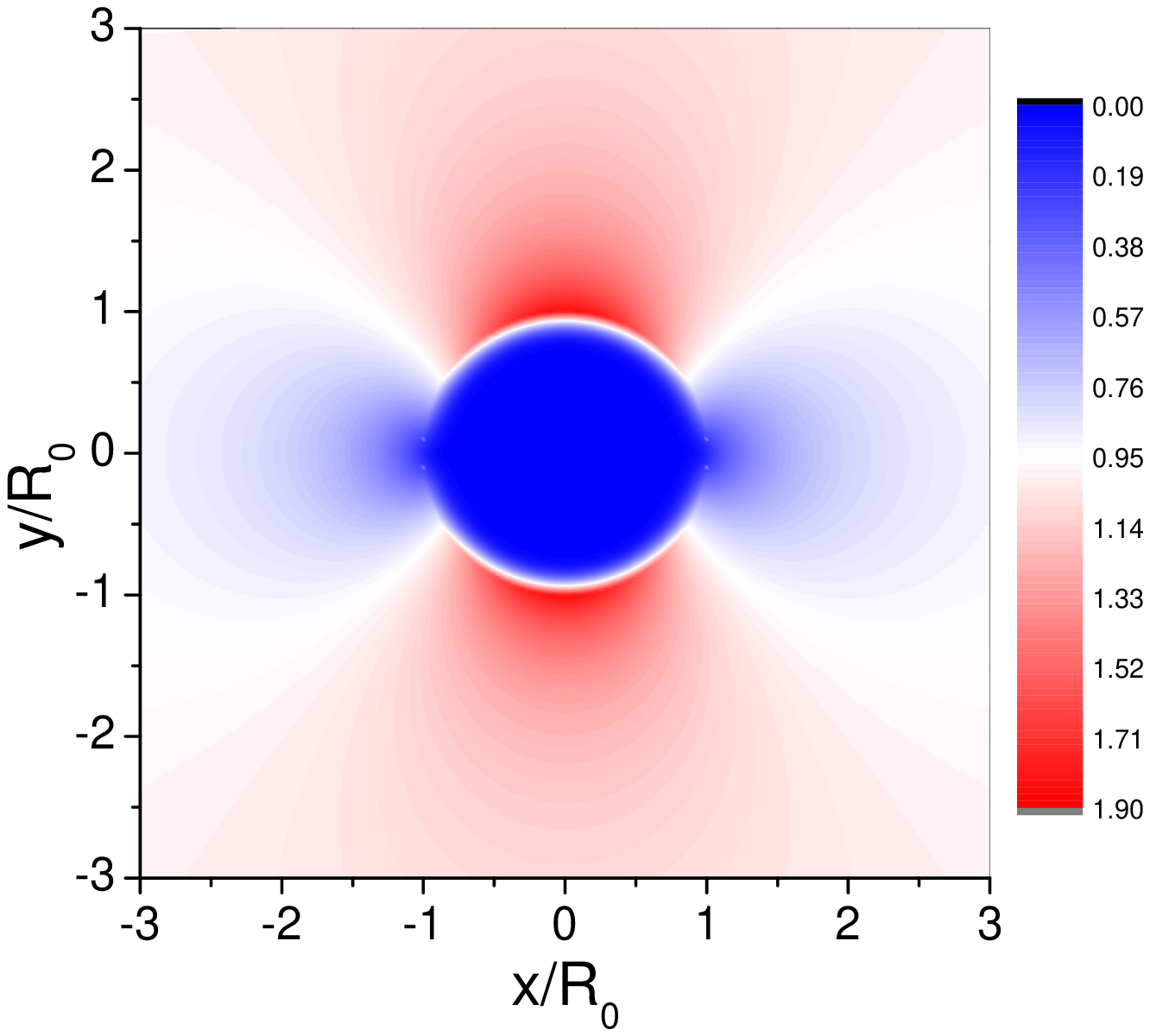}
\end{center}
\caption{(Color online) The distribution of the magnetic field strength (in units of $H_{0}$) in the $xy$
plane at $t =0.1 \tau_{\mathrm{R}}$ for $\delta =1$ and for $\theta =0$ (left panel), $\theta =\pi /4$ (middle panel),
and $\theta =\pi /2$ (right panel). The coordinates are scaled by $R_{0}$.}
\label{fig:4}
\end{figure*}

\begin{figure*}[tbp]
\begin{center}
\includegraphics[width=55.0mm]{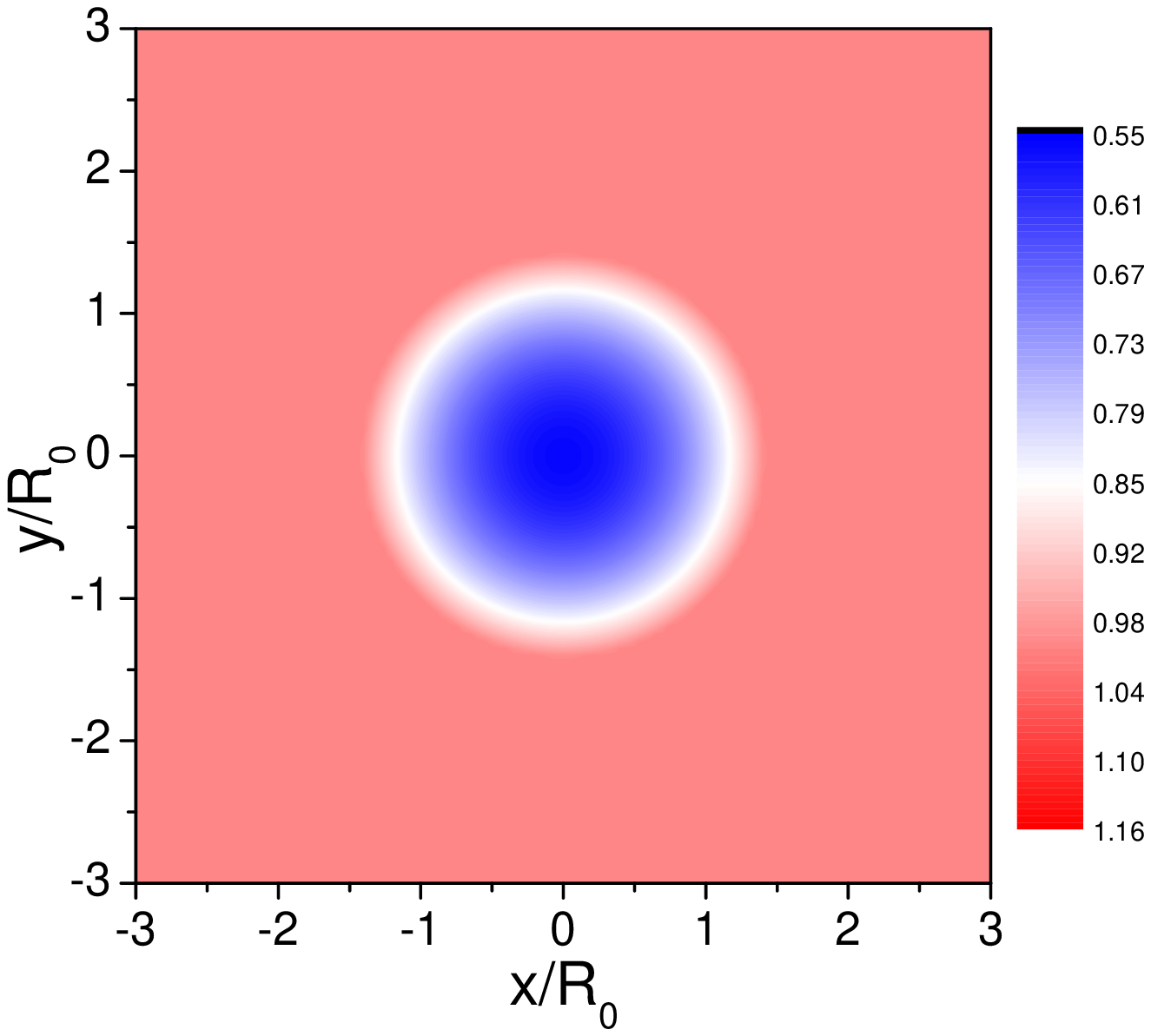} %
\includegraphics[width=55.0mm]{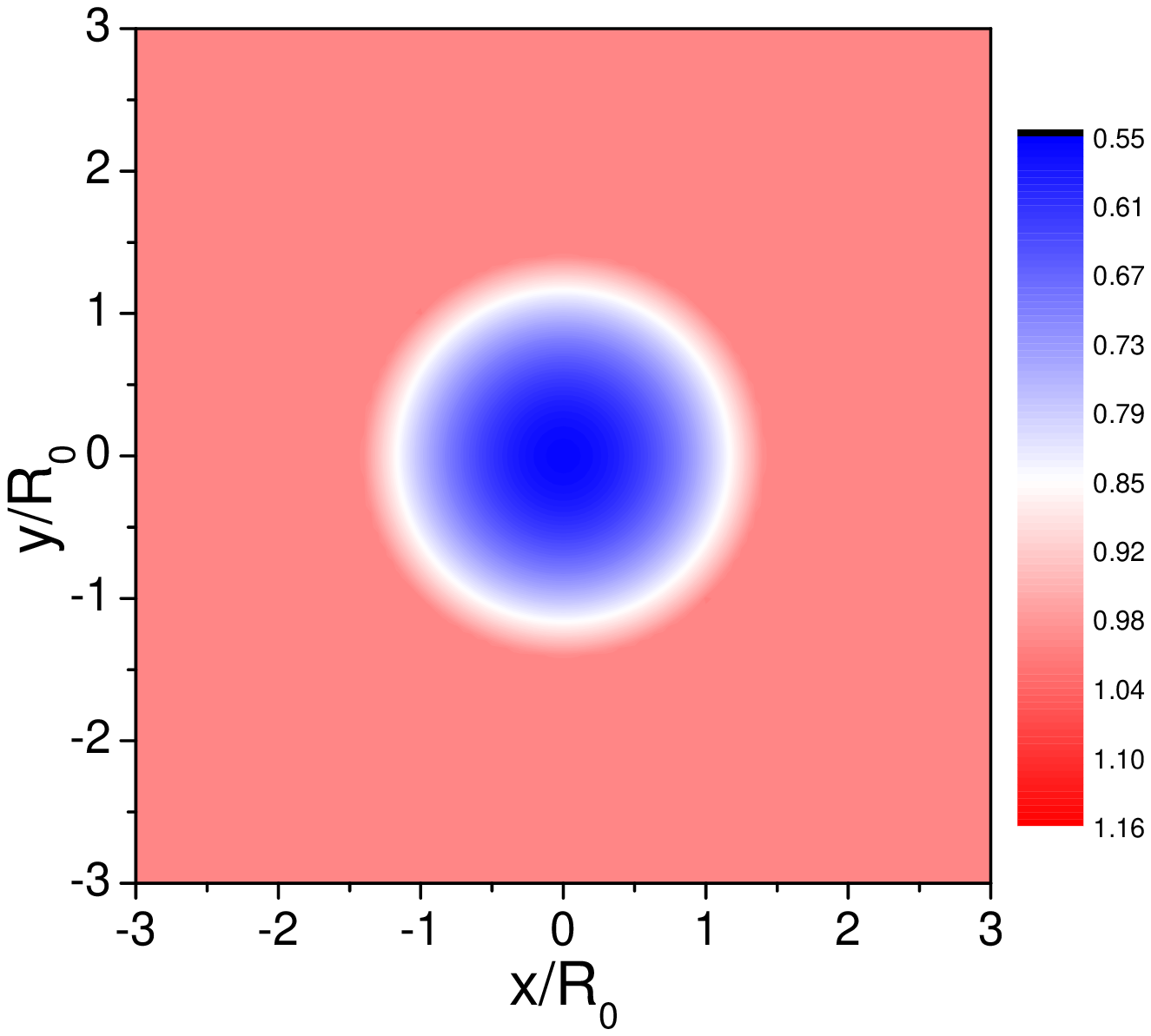} %
\includegraphics[width=55.0mm]{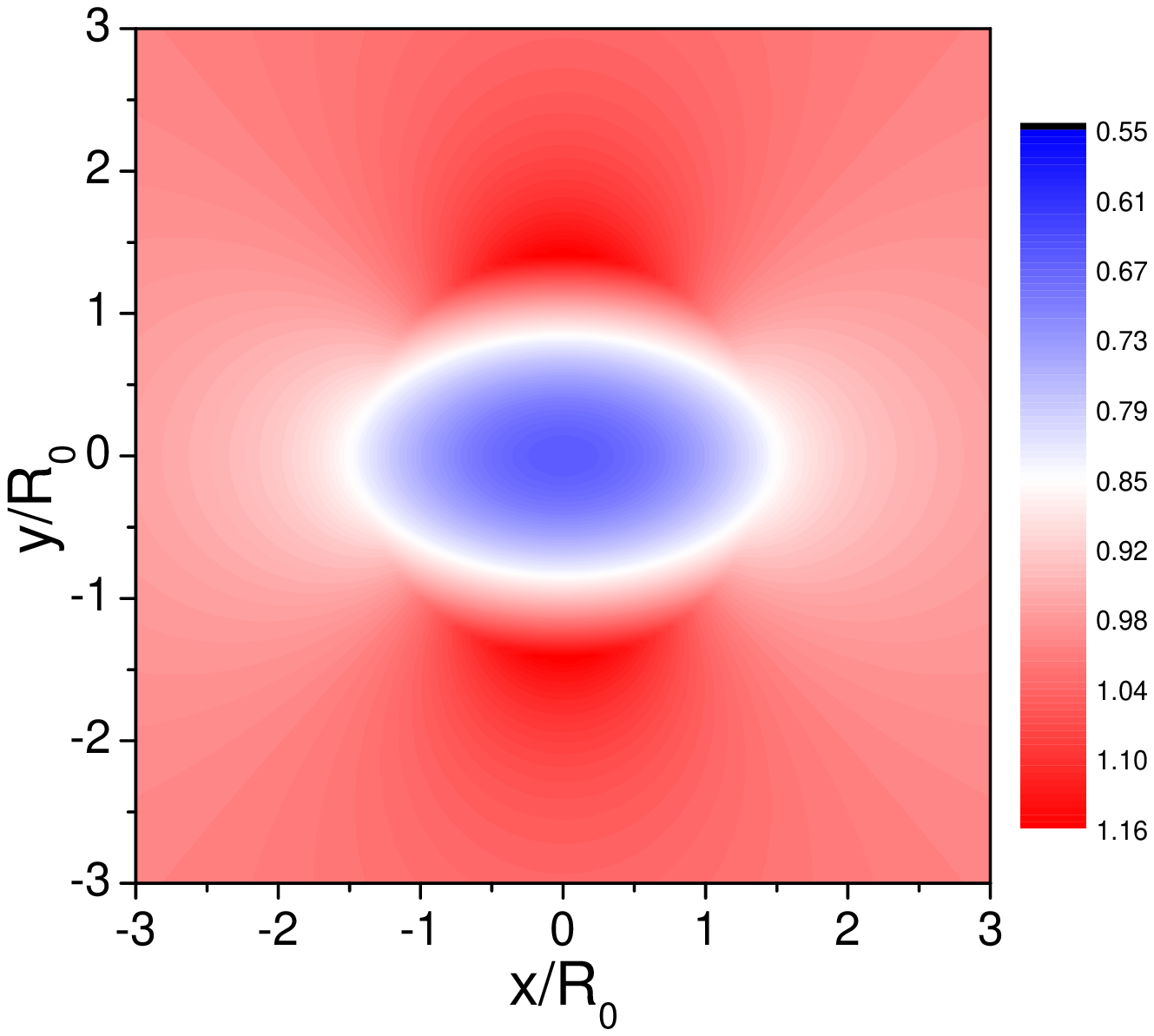}
\end{center}
\caption{(Color online) Same as in Fig.~\ref{fig:4} but at $t= \tau_{\mathrm{R}}$.}
\label{fig:5}
\end{figure*}

\begin{figure*}[tbp]
\begin{center}
\includegraphics[width=55.0mm]{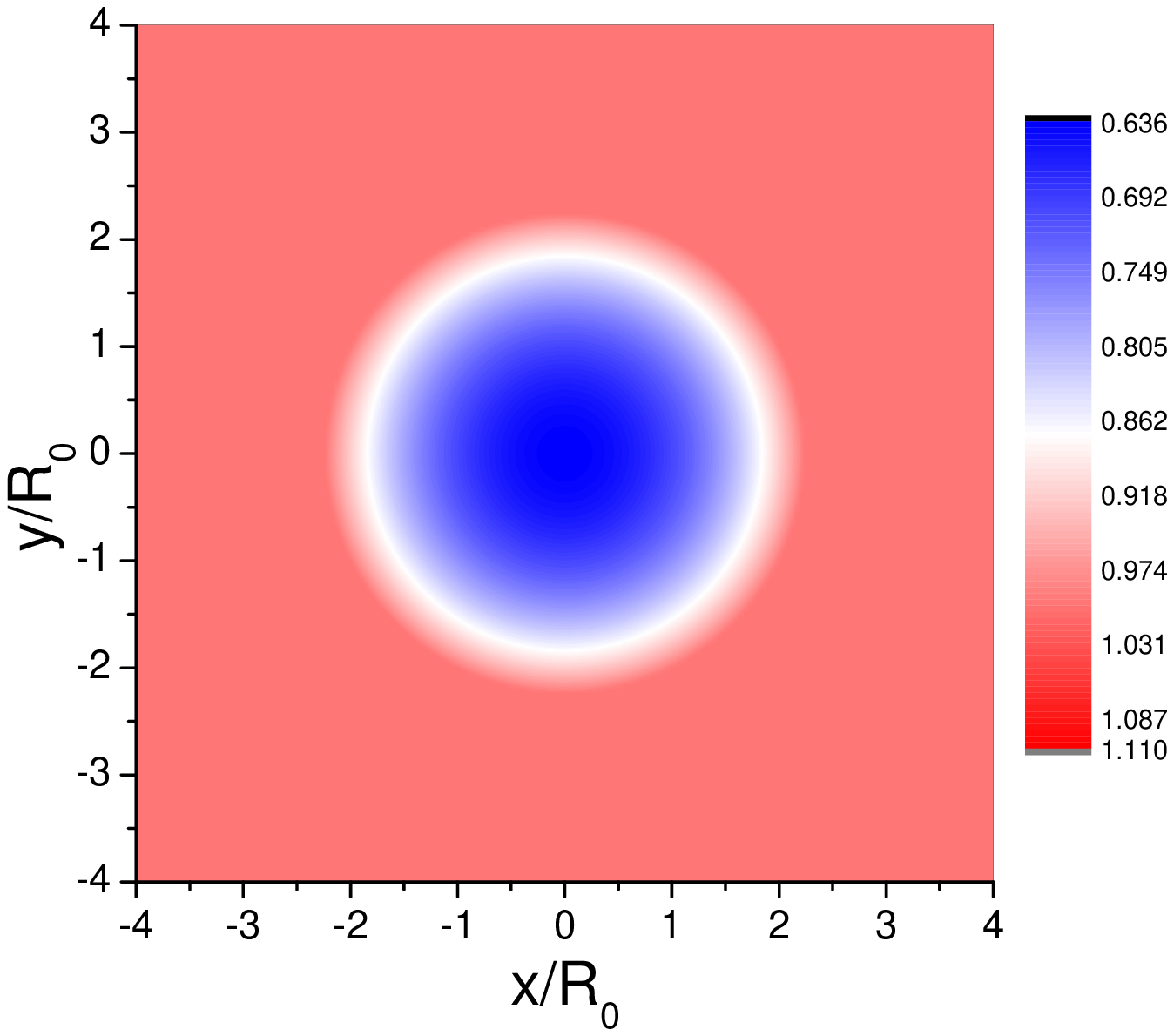} %
\includegraphics[width=55.0mm]{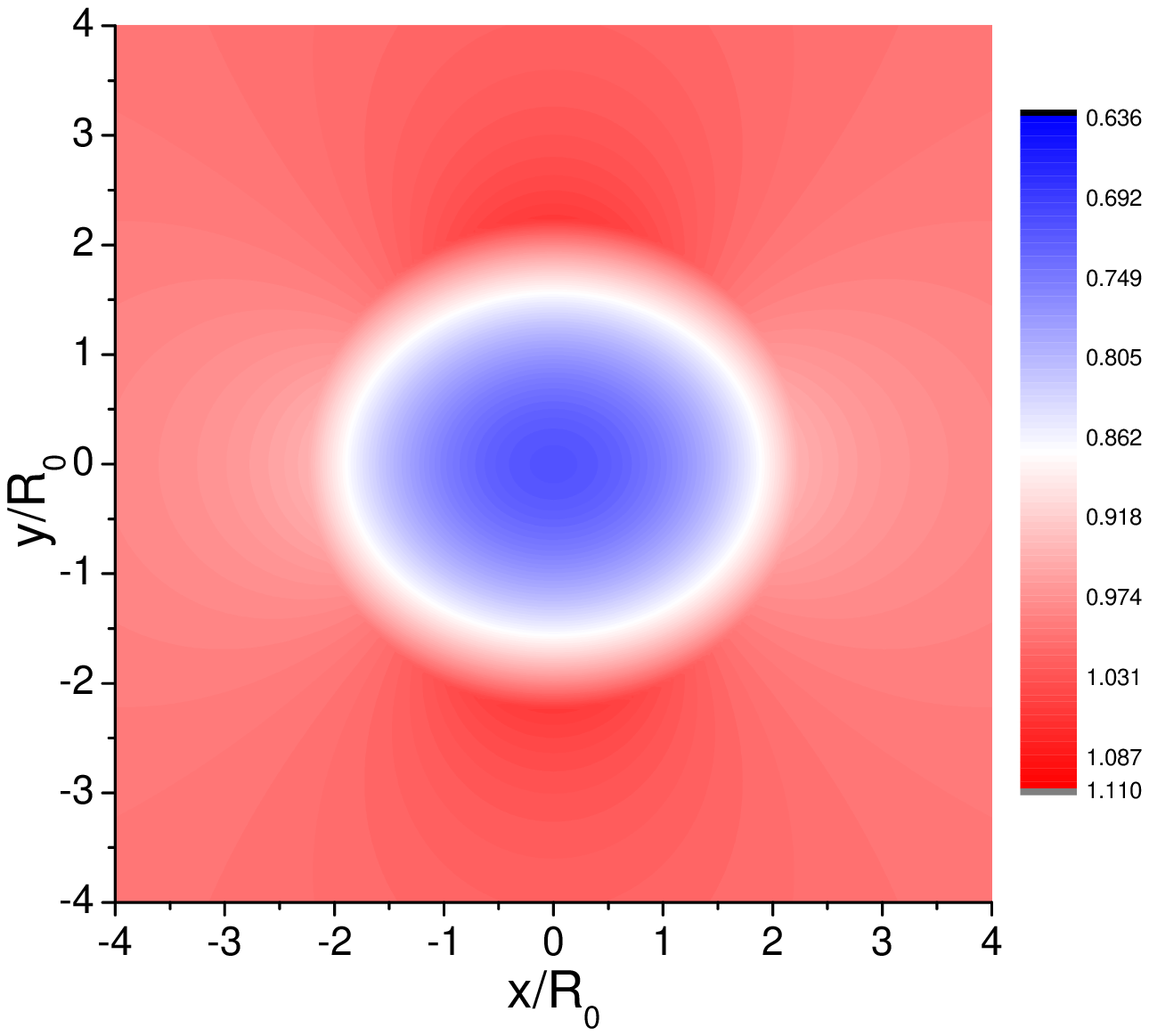} %
\includegraphics[width=55.0mm]{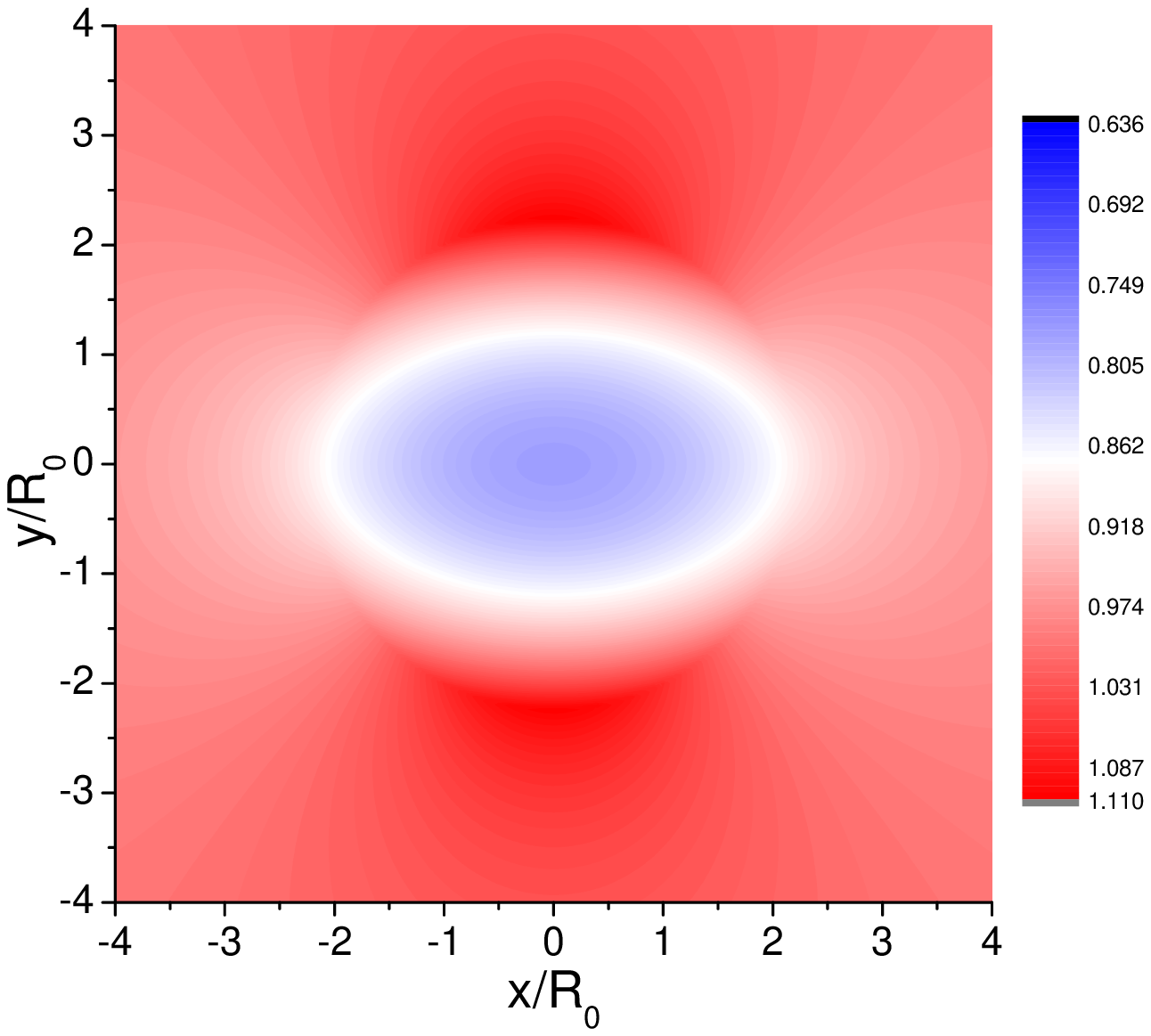}
\end{center}
\caption{(Color online) Same as in Fig.~\ref{fig:4} but at $t= 2\tau_{\mathrm{R}}$.}
\label{fig:6}
\end{figure*}

Next, we consider the temporal and spatial distributions of the magnetic field for various diffusion
coefficients $\delta $ and for orientation angles $\theta $. Figure~\ref{fig:3} demonstrates the absolute
values of the magnetic field $H(r,t)$ (in units of $H_{0}$) on the symmetry axis of the plasma cylinder,
$r=0$, as a function of time. As it follows from Eqs.~\eqref{eq:28}--\eqref{eq:32} the electric field
vanishes on this axis, $\mathbf{E} =0$, while $H_{r}=H_{0\perp }\mathcal{F}_{1}(t)\cos \varphi $,
$H_{\varphi }=-H_{0\perp }\mathcal{F}_{1}(t)\sin \varphi $, $H_{z}=H_{0\parallel }\mathcal{F}_{2}(t)$, where
\begin{eqnarray}
&&\mathcal{F}_{1}(t)=1-\frac{2}{R(t)}\sum_{n=1}^{\infty }\frac{T_{n}(t)}{%
\lambda _{n}J_{1}(\lambda _{n})} ,  \label{eq:53} \\
&&\mathcal{F}_{2}(t)=1-\frac{2}{R^{2}(t)}\sum_{n=1}^{\infty }\frac{U_{n}(t)}{%
\lambda _{n}J_{1}(\lambda _{n})} .  \label{eq:54}
\end{eqnarray}%
Thus, as expected, the magnetic field $H(r,t)$ is independent on the angle $\varphi $. From Fig.~\ref{fig:3}
it is seen that simultaneously with the plasma expansion the magnetic field inside grows monotonically from
zero value and saturates typically at $t_{\mathrm{sat}}\simeq R_{0}^{2}/D_{0}\sim \delta ^{-1}$. Clearly the
saturation time of the magnetic field decreases with $\delta$. At $t>t_{\mathrm{sat}}$ the magnetic field is
constant because it is completely diffused into the expanding plasma remaining, however, smaller than $H_{0}$.
In this asymptotic regime from Eqs.~\eqref{eq:49} and \eqref{eq:50} one obtains $T_{n}(t)\simeq R(t)/(\lambda%
_{n}^{2}\delta +1)$ and $U_{n}(t)\simeq R^{2}(t)/(\lambda _{n}^{2}\delta /2+1)$. Insertion of these expressions
into Eqs.~\eqref{eq:53} and \eqref{eq:54} determines the magnetic field at $t>t_{\mathrm{sat}}$. It is also
noteworthy the influence of the orientation of $\mathbf{H}_{0}$ on the magnetic field inside the plasma cylinder
at $r=0$. At weak diffusion, $\delta \ll 1$, the quantities $\mathcal{F}_{1}(t)$ and $\mathcal{F}_{2}(t)$
are evaluated using Eq.~\eqref{eq:ap10} with $x=0$ and $z= \delta^{-1/2}$. In this case we obtain $\mathcal{F}_{1}(t)%
\simeq (4\pi ^{2}/\delta )^{1/4} \exp (-1/\sqrt{\delta })$ and $\mathcal{F}_{2}(t)\simeq (8\pi ^{2}/\delta )^{1/4}
\exp (-\sqrt{2/\delta })$. It is seen that $\mathcal{F}_{1}(t)\gg \mathcal{F}_{2}(t)$ and the magnetic field is essentially
larger for the transverse orientation of $\mathbf{H}_{0}$ (Fig.~\ref{fig:3}, left panel). However, in the strongly
diffusive regime, $\delta \gg 1$, from Eqs.~\eqref{eq:53} and \eqref{eq:54} one derives $\mathcal{F}_{1}(t)\simeq 1-%
1/4\delta $ and $\mathcal{F}_{2}(t)\simeq 1-1/2\delta $, and the magnetic field is only weakly sensitive to
$\theta$ approaching the value $H_{0}$ of the unperturbed magnetic field (Fig.~\ref{fig:3}, right panel).

More complete spatio-temporal evolution of the magnetic field is shown in Figs.~\ref{fig:4}--\ref{fig:6}. Shown are
the distributions of the magnetic field strength in the $xy$-plane transverse to the symmetry axis of the plasma
cylinder ($z$-axis) at $t= 0.1\tau_{\mathrm{R}}$ (Fig.~\ref{fig:4}), $t= \tau_{\mathrm{R}}$ (Fig.~\ref{fig:5}), and $t= 2\tau_{\mathrm{R}}$
(Fig.~\ref{fig:6}) and for $\delta =1$, $\theta =0$, $\theta =\pi /4$, and $\theta =\pi /2$. Note that the plasma
radii are $R(t) \simeq 1.005R_{0}$, $R(t) \simeq 1.4R_{0}$, and $R(t) \simeq 2.2R_{0}$ in Figs.~\ref{fig:4},
\ref{fig:5}, and \ref{fig:6}, respectively. It is seen that the magnetic field distribution essentially depends on
the orientation angle $\theta$ of the unperturbed magnetic field $\mathbf{H}_{0}$. At the initial stage of the
plasma expansion the magnetic field inside is excluded for all $\theta$ except the domains in the close vicinity
of the plasma boundary ($r\simeq R(t)$) where the magnetic field diffuses into the plasma due to the finite
conductivity. Moreover, outside the expanding plasma and at parallel orientation with $\theta =0$ the magnetic field
is not perturbed ($\mathbf{H} =\mathbf{H}_{0}$) while at the oblique ($\theta =\pi /4$) and transverse ($\theta =\pi /2$)
orientations the magnetic field distribution exhibits much more complex behavior. It is clearly seen that the expanding
plasma pushes the magnetic field out in the direction of the transverse component $\mathbf{H}_{0\perp}$ (i.e. at
$\varphi =0$ or $\varphi =\pi$) where the strength of the magnetic field exceeds essentially the unperturbed value
$H_{0}$. This effect is maximal at $\theta =\pi /2$ when $H\simeq 1.9H_{0}$ (see Fig.~\ref{fig:4}, right panel) in
the vicinity of the plasma boundary at $\varphi =0$ and $\varphi =\pi $. In the plane transverse to $\mathbf{H}_{0\perp}$
and containing the $z$--axis (i.e. at $\varphi =\pi /2$ and $\varphi =3\pi /2$) the expanding plasma forms a magnetic
cavity outside its volume where the strength of the magnetic field is reduced and $H\simeq 0.8H_{0}$ (see Fig.~\ref{fig:4},
middle and right panels). Thus, at this early stage of the expansion of the highly conducting plasma the role of the
magnetic field diffusion is small.

Next, at the later time, $t\gtrsim \tau_{\mathrm{R}}$, the electrical conductivity of the plasma is reduced and the
magnetic field is essentially diffused into the plasma which is more effective at larger $\theta$ (Figs.~\ref{fig:5}
and \ref{fig:6}). Furthermore, in the domain outside the expanding plasma volume the diffusion smooths out the magnetic
field distribution and it is only weakly perturbed for parallel and oblique orientations of $\mathbf{H}_{0}$. At transverse
orientation, however, the magnetic field deviates considerably from $\mathbf{H}_{0}$. In particular, it is noteworthy
that in the latter case the magnetic cavity is compressed along the direction of $\mathbf{H}_{0\perp}$ ($y$--axis).
Finally, at $t\gtrsim 2\tau_{\mathrm{R}}$ with a further reduction of the plasma conductivity the magnetic field diffuses
considerably into the plasma (Fig.~\ref{fig:6}) eventually recovering the uniform distribution in the $xy$-plane. However,
the strength of the magnetic field inside the plasma volume is systematically smaller than $H_{0}$ as shown in Fig.~\ref{fig:3}.

\section{Conclusion}
\label{sec:6}

In this paper, an exact solution of the purely radial expansion of a neutral, resistive plasma cylinder in the presence
of a magnetic field has been obtained. The electromagnetic fields are derived by using the appropriate initial and
boundary conditions. Two kind of the initial conditions have been considered assuming poorly or perfectly conducting
plasma at the initial state. In the first case the external magnetic field is completely penetrated inside the plasma
while in the other extreme case the magnetic field is completely excluded from the initial plasma volume. However, as
expected both solutions "forget" the imposed initial conditions at $t\gtrsim \tau_{\mathrm{D}}$, where $\tau_{\mathrm{D}}$
is the characteristic diffusion time.
Using a simple model for the electrical conductivity we have also studied the energy balance during the plasma expansion
as well as the spatio-temporal distribution of the magnetic field with an arbitrary orientation of the initial field.
In contrast to the previous treatments with highly conducting plasmas \cite{kat61,rai63,dit00,ner06,ner12,ner10}, our
model calculations demonstrate some new features arising due to the finite conductivity of the expanding plasma.

Going beyond the present model, which is based on several approximations, we can envisage a number of improvements.
We have assumed the purely radial expansion (with a given velocity $\dot{R} (t)$ of the plasma boundary) and hence the
shape of the plasma (in the $xy$-plane) remains isotropic during plasma expansion. For realistic laser-generated plasmas
this is valid when the kinetic energy density of the expanding plasma much exceeds the magnetic field energy density \cite{zak03,win05}.
However, after some period of accelerated motion, the plasma gets decelerated as a result of the external Lorentz force
acting inward and the above mentioned condition may be violated at the later stages ($t\gg \tau_{\mathrm{D}}$) of the
expansion. In this case the Lorentz force density which is
anisotropic in general, cannot be neglected in Eq.~\eqref{eq:a1}. Thus one can expect some deformation of the initially
isotropic plasma surface \cite{osi03,ner09} and the plasma radius $R(t,\varphi )$ should be treated as a function of
$\varphi$. On the other hand, in this regime the plasma radius $R(t,\varphi )$ cannot be treated as a given function of
time and should be determined self-consistently using, for instance, the equation of balance of plasma ($\sim \rho \dot{R}^{2}$)
and magnetic field ($\sim H^{2}$) energies \cite{ner11}. Thus, plasma dynamics can be described (at least qualitatively)
inserting Eqs.~\eqref{eq:28}-\eqref{eq:30} and \eqref{eq:33}, \eqref{eq:34} into the energy balance equation which yields
a first-order differential equation for $R(t,\varphi )$. Within the scope of this analysis, the deformation of the plasma
surface as well as the initial stage of plasma acceleration, the later stage of deceleration and the process of stopping
at the point of maximum expansion could be examined numerically. We intend to address these issues in our forthcoming
investigations.

\begin{acknowledgments}
This work has been supported by the State Committee of Science of Armenian
Ministry of Higher Education and Science (Project No.\ 11--1c317).
\end{acknowledgments}

\appendix

\section{Some summation formulas involving the Bessel functions}
\label{sec:app1}

Using the Fourier-Bessel expansion of an arbitrary function of a real
variable \cite{bat53} one can derive some summation formulas involving the
Bessel functions which are used in the main text of the paper. The first
relation is obtained by considering the Fourier-Bessel series for the
quadratic function $x^{2}$ in the interval $0\leqslant x<1$,
\begin{equation}
x^{2}=\sum_{n=1}^{\infty } \left( 1-\frac{4}{\lambda _{n}^{2}}\right)
\frac{2 J_{0}(\lambda _{n}x)}{\lambda _{n}J_{1}(\lambda _{n})} ,
\label{eq:ap1}
\end{equation}%
where $\lambda _{n}$ (with $n=1,2,...$) are the positive zeros of the Bessel
function, $J_{0}(\lambda _{n})=0$, arranged in ascending order of magnitude \cite{bat53}.

On the other hand using the known summation formula \cite{bat53} (valid at $0\leqslant x<1$)
\begin{equation}
\sum_{n=1}^{\infty }\frac{J_{0}(\lambda _{n}x)}{\lambda _{n}J_{1}(\lambda_{n})}=\frac{1}{2}
\label{eq:ap4}
\end{equation}%
one can represent Eq.~\eqref{eq:ap1} in another form
\begin{equation}
\sum_{n=1}^{\infty }\frac{J_{0}(\lambda _{n}x)}{\lambda_{n}^{3}J_{1}(\lambda _{n})}=\frac{1-x^{2}}{8} .
\label{eq:ap5}
\end{equation}%
This latter relation is valid at $0\leqslant x\leqslant 1$.

Next taking the $x$-derivatives of Eqs.~\eqref{eq:ap1} and \eqref{eq:ap4}
one obtains
\begin{equation}
\sum_{n=1}^{\infty }\left( 1-\frac{4}{\lambda _{n}^{2}}\right) \frac{%
J_{1}(\lambda _{n}x)}{J_{1}(\lambda _{n})}=-x, \,\,
\sum_{n=1}^{\infty }\frac{J_{1}(\lambda _{n}x)}{J_{1}(\lambda _{n})}=0 ,
\label{eq:ap6}
\end{equation}%
respectively. The latter formula is valid at $0\leqslant x<1$. Therefore, using the second relation in
Eq.~\eqref{eq:ap6} the first summation formula in \eqref{eq:ap6} can be represented in the form
\begin{equation}
\sum_{n=1}^{\infty }\frac{J_{1}(\lambda _{n}x)}{\lambda
_{n}^{2}J_{1}(\lambda _{n})}=\frac{x}{4} .
\label{eq:ap8}
\end{equation}

Finally, substituting $x=1$ in Eq.~\eqref{eq:ap8} we arrive at
\begin{equation}
\sum_{n=1}^{\infty }\frac{1}{\lambda _{n}^{2}}=\frac{1}{4} .
\label{eq:ap9}
\end{equation}

Next we derive another kind of summation formulas involving the zeros of the
Bessel function $J_{0}(z)$. For that purpose consider the known summation
formula \cite{bat53}
\begin{equation}
\frac{I_{0}(xz)}{I_{0}(z)}=2\sum_{n=1}^{\infty }\frac{\lambda
_{n}J_{0}(\lambda _{n}x)}{\left( \lambda _{n}^{2}+z^{2}\right) J_{1}(\lambda _{n})} ,
\label{eq:ap10}
\end{equation}%
where $I_{n}(z)$ (with $n=1,2,...$) is the modified Bessel function of the first kind, $z$ and $x$
(with $0\leqslant x<1$) are real variables. Differentiating Eq.~\eqref{eq:ap10} with respect to
$x$ and using the second relation in Eq.~\eqref{eq:ap6} we arrive at
\begin{equation}
\frac{I_{1}(xz)}{2zI_{0}(z)}=\sum_{n=1}^{\infty }\frac{J_{1}(\lambda _{n}x)}{%
\left( \lambda _{n}^{2}+z^{2}\right) J_{1}(\lambda _{n})}
\label{eq:ap11}
\end{equation}%
which, in particular, at $x=1$ yields
\begin{equation}
\sum_{n=1}^{\infty }\frac{1}{\lambda _{n}^{2}+z^{2}}=\frac{I_{1}(z)}{2zI_{0}(z)} .
\label{eq:ap12}
\end{equation}%
This relation can be used for deriving other summation formulas. For
instance, extracting Eq.~\eqref{eq:ap12} from Eq.~\eqref{eq:ap9} and after
some manipulations one obtains
\begin{eqnarray}
&&\Xi _{1}(z)=\sum_{n=1}^{\infty }\frac{4}{\lambda _{n}^{2} ( \lambda
_{n}^{2} z+1) } = 1- \frac{2I_{1} (\zeta )}{\zeta I_{0} (\zeta )} ,  \label{eq:ap13} \\
&&\Xi _{2}(z)=\sum_{n=1}^{\infty }\frac{4}{\lambda _{n}^{2} (\lambda_{n}^{2} z+1) ^{2}}
=2-\frac{4I_{1} (\zeta )}{\zeta I_{0} (\zeta )} -
\left[\frac{I_{1}(\zeta )}{I_{0}(\zeta )}\right] ^{2} .  \label{eq:ap14}
\end{eqnarray}%
Here $\zeta =1/\sqrt{z}$. We also mention the asymptotic behavior of the
functions $\Xi _{1}(z)$ and $\Xi _{2}(z)$. At small argument, $z\ll 1$,
these functions behave as $\Xi _{1}(z)\simeq \Xi _{2}(z)\simeq 1$ while at $z\gg 1$
they decay as $\Xi _{1}(z)\simeq 1/8z$, $\Xi _{2}(z)\simeq 1/48z^{2}$.

\end{document}